\documentclass[12pt,preprint]{aastex}
\usepackage{graphicx}
\usepackage{color}

\newcommand {\go} {\mathrel{\hbox{\rlap{\lower.55ex \hbox {$\sim$}}
        \kern-.3em \raise.4ex \hbox{$>$}}}}
\newcommand {\lo} {\mathrel{\hbox{\rlap{\lower.55ex \hbox {$\sim$}}
        \kern-.3em \raise.4ex \hbox{$<$}}}}
\newcommand {\be} {\begin{equation}}
\newcommand {\ee} {\end{equation}}

\newcommand {\taup} {\tau_{\rm p}}
\newcommand {\lh} {l_{\rm h}}
\newcommand {\lsoft} {l_{\rm s}}
\newcommand {\kT} {k T_{\rm e}}

\shorttitle{High-$z$ RQQs: patchy corona model.}

\shortauthors{Sobolewska  et al.}

\begin{document}

\title{High redshift radio quiet quasars -- exploring parameter space 
of accretion models. Part II: patchy corona model} 
\author{Ma{\l}gorzata A. Sobolewska}
\affil{Harvard-Smithsonian Center for Astrophysics, 60 Garden Street, Cambridge, MA 02138\\
Nicolaus Copernicus Astronomical Center, Bartycka 18, 00-716 Warszawa, Poland}
\email{msobolewska@cfa.harvard.edu, malsob@camk.edu.pl}

\author{Aneta Siemiginowska}
\affil{Harvard-Smithsonian Center for Astrophysics, 60 Garden Street, Cambridge, MA 02138}
\email{asiemiginowska@cfa.harvard.edu}

\and

\author{Piotr T. \.{Z}ycki} 
\affil{Nicolaus Copernicus Astronomical Center, Bartycka 18, 00-716 Warszawa, Poland}
\email{ptz@camk.edu.pl}

\begin{abstract} 
We modeled the spectral energy distribution (SED) of high redshift
radio quiet quasars (high-$z$ RQQs). We computed spectra in a
patchy corona geometry where an accretion 
disk extends to the last stable orbit and the Comptonizing active
regions (hot clouds) are distributed above the disk. We explored the model
parameter space to find theoretical parameters that give spectra
with the optical/UV luminosity, the X-ray loudness, and the X-ray
photon index compatible with those of high-$z$ RQQs observed with 
{\it Chandra}. We found that a range of solutions is possible, from
high-$kT_{\rm e}$ low-$\tau$ to low-$kT_{\rm e}$ high-$\tau$
models. The solutions require low level of energy dissipation in the
hot clouds and low disk covering factor. The modeled mass is of the order of
$10^{10}\, M_{\odot}$ and the accretion rate is $\dot{M} \ge 0.2\, \dot{M}_{\rm
  Edd}$. We compare our results to those obtained previously for hot
inner flow geometry. 
\end{abstract}


\section{Introduction}

Modeling quasars' broad band energy spectra provides important
information about the geometry and environment of the quasars at
different epochs.
Such spectra are thought to originate in an accretion
of matter onto a massive black hole (Rees 1984). In the
standard scenario a thin accretion disk that is formed thermalizes
the gravitational potential energy and radiates it as a disk black
body giving a characteristic bump in the spectrum (Shakura \& Sunyaev
1973). In the case of high redshift quasars, this bump is shifted into
the optical/UV range (Shields 1978), which makes it easy to
observe and study. The X-ray part of the spectrum is believed to
originate in an inverse Compton
process of the disk photons on energetic electrons. Modeling the X-ray
component gives a direct information about the geometry of the
accretion flow (i.e.\ the relative location of the cold accretion disk
and Comptonizing plasma).  Radio quiet quasars are good objects for
such studies since their X-ray spectra are not ``contaminated'' by any
additional components connected to the radio emission.
Moreover, the radio quiet quasars comprise majority among the quasars
population (Stern et al.\ 2000). They are more common, and hence more
typical. Recent X-ray observations of quasars located at redshift
$z>4$ (e.g. Becker et al.\ 2001; Brandt et al.\ 2002a; Bechtold et al.\
2003, hereafter B03) provide for the first time a unique opportunity
to study these objects at the early Universe and investigate their
evolution.  Thus, we focus our study on the high redshift radio quiet
quasars (high-$z$ RQQs).

The observations of accreting sources support the hypothesis that the
accretion  flow is composed of a cold medium producing a thermal
radiation and a hot medium radiating in X-rays. The puzzling question
is that of a relative location of these two phases. A variety of
geometries has been considered in order to explain the observed
spectra and correlations. In Sobolewska, Siemiginowska \& \.{Z}ycki
2004 (hereafter Paper I) we shortly reviewed currently considered
scenarios for the accretion flow which were successful in explaining
spectra of galactic X-ray binaries (XRB) and Seyfert galaxies. We
discussed possible tests of geometry that can be conducted based on
modeling the broad band spectra. In particular, a correlation between
the amplitude of reflection, $R$, and the X-ray photon index,
$\Gamma$, observed in X-ray binaries and Seyfert galaxies (Zdziarski,
Lubi\'nski, \& Smith 1999) seems to be one of the criteria. Such
correlations can be 
reproduced by the following models: (1) a truncated disk surrounding
a hot inner flow (Zdziarski et al.\ 1999), (2) a hot corona formed by
active regions  above the disk, outflowing at relativistic speeds
(Beloborodov 1999a, hereafter B99), (3) a cold disk covered by a hot
ionized skin which decreases the effectiveness of thermalization
(Nayakshin, Kazanas \& Kallman 2000) (4) a hot spherical inner flow
surrounded by cold clouds (Malzac 2001). In Paper I 
we investigated in
details the first possibility. We assumed that the accretion flow in
high-$z$ RQQs proceeds through a standard optically thick cold
accretion disk covered with a hot corona. At a certain radius the disk
was assumed to evaporate and a hot inner flow was formed. We computed
spectra in such geometry and explored the model parameter space. We
found that modeling both the optical/UV and X-ray component provides
tight constraints on the Comptonizing plasma parameters through the
observed X-ray loudness parameter, $\alpha_{\rm ox}$, X-ray photon
index, $\Gamma$, and the optical/UV luminosity, $l_{\rm UV} \equiv
\log(\nu{L_{\nu}})$ at $\nu = c/{\lambda}_{\rm 2500\, \AA}$ in the
rest frame. For most objects the truncation radius is required
to be larger than the radius of the innermost stable orbit, 
$r_{\rm tr}>3\,R_{\rm S}$, with an upper limit 
of $r_{\rm tr} \leq 40\, R_{\rm S}$ (where $R_{\rm S} =
\frac{2GM}{c^2}$ is a Schwarzschild radius). We also concluded that the
advection in the inner hot flow may play an important role. Moreover,
the shape of the inner hot flow may deviate from spherical
one. Instead, it may be considerably flattened. Our modeling
suggested high accretion rates, $\dot{M} \ge 20\% \dot{M}_{\rm  Edd}$
which may suggest similarity of high-$z$ RQQs to XRBs in the high
state (HS) or very high state (VHS) of black hole X-ray binaries. 
However, in these spectral states the 
accretion disk is thought to extend to the last stable orbit.
This provides motivation for investigating the geometry of a disk
extending to the innermost stable orbit, with some kind of an X-ray active
corona.

In this paper we investigate such an active corona geometry.
The structure of
the paper is as follows. In Section~\ref{sec:model} we describe the
model. In Section 3. we present the results. Section 4. contains
application of the model to the sample of high-$z$ quasars from
B03. Finally, in Section 5. we discuss the results and give the
concluding remarks.

\section{Model}
\label{sec:model}

\begin{figure*}
\epsscale{0.8}
\plotone{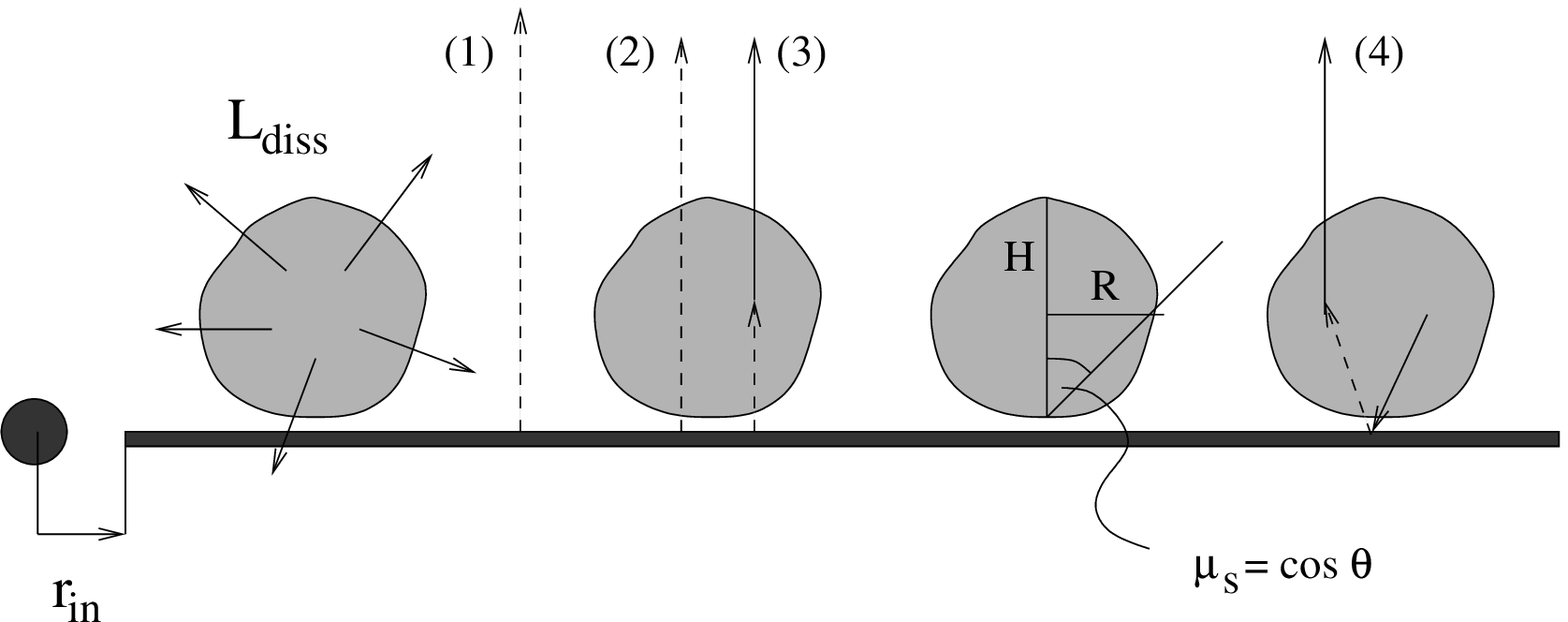}
\caption{Patchy corona geometry. The accretion disk extends to the
last stable orbit, $r_{\rm in}$. The disk is covered with a hot clouds
whose total luminosity accounts for $L_{\rm diss}$. The spectral components
are: the disk blackbody-like radiation which escapes the system without
encountering hot plasma (1) or without being scattered in the hot plasma
(2), and Comptonized component. The soft photons for Comptonization come
from viscous dissipation in the disk (3) or reprocessing of the hot plasma
radiation (4). The interpretation of the clouds geometry parameter,
$\mu_{\rm s}$, is illustrated. See also Figure \ref{fig:mus}.}
\label{fig:geom}
\end{figure*}

The plane-parallel
geometry in which the accretion disk extends to the innermost circular orbit and
the hot corona is located above the disk does not reproduce the photon
index less than $\sim$2 (Haardt \& Maraschi 1991, Stern et al.\ 1995). According to Vignali et 
al.\ (2003a, hereafter V03) the mean X-ray photon index
of high-$z$ radio quiet 
quasars is $\Gamma \sim 2$, so the model in which this
value is somehow limiting does not seem to offer enough flexibility in
explaining the distribution of photon indices around the derived
mean. Moreover, the mean X-ray photon index from the
sample of B03 is $\sim 1.5$. The
plane parallel geometry can be modified in order to account for harder
spectra. Here, we consider a case in which the hot
Comptonizing plasma is not continuous but is composed of a number of active regions
distributed above the accretion disk
(driven by, e.g., magnetic reconnection events; Galeev, Rosner, \& Vaiana
1979; Haardt, Maraschi, \& Ghisellini 1994; B99; Malzac,
Beloborodov, \& Poutanen 2001). The geometry is schematically presented in
Figure 
\ref{fig:geom}. Additionally, the hot plasma can move toward to or away
from the disk (B99). 

We assume that the innermost stable orbit is located at $r_{\rm in} =
3\, R_{\rm S}$, and the accretion efficiency is $\epsilon=1/12$. 

The computed spectra are a sum of a Comptonized component and  thermal disk
radiation that escapes the hot plasma without being scattered or does not
encounter the hot plasma on its way to the observer.

\subsection{The Comptonized component}

We consider a thermal Comptonization case, i.e.\ we assume that the Comptonizing 
electrons have purely thermal (Maxwellian) energy distribution.
We compute the Comptonized component using the {\sc EQPAIR\/} code of P.\ Coppi,
described in detail in, e.g., Coppi (1999). The main parameters of the code
are the plasma heating rate, $\lh$, and the soft photon compactness,
$\lsoft$ ($l \equiv (L/R) \sigma_{\rm T} /(m_{\rm e} c^3)$, where $L$ is 
the luminosity and $R$ is the size of the plasma cloud). In addition, we need to specify
the spectrum of input soft photons, and background plasma optical depth, $\taup$.
The code includes all relevant microphysical processes of interactions between
plasma and radiation. It computes plasma parameters: the temperature, $\kT$ and total 
optical depth, including the contribution from electron-positron pairs,
as well as the spectrum of Comptonized emission.

In practice we parameterize our results by the ratio $\lh/\lsoft$,
and the plasma temperature, $\kT$. The former is convenient since it is easily
computed from an assumed geometry (see below), and it is the most important parameter
to determine the spectral slope of radiation (e.g.\ Beloborodov 1999b).
The temperature replaces  the heating compactness, $\lh$, which is more difficult to estimate
observationally. To achive the required value of $\kT$ one should
iterate the computations over the total optical depth. In practise,
we iterate over $\lh$ for a pure pair plasma (see also Haardt \& Maraschi
1991). We consider the low- and high- temperature regimes ($kT_{\rm
  e}=100$ and 300 keV, respectively), since there are 
no observational data to constrain this parameter in the case of high-$z$ RQQs.
The spectrum of soft input photons is the disk blackbody
spectrum constructed as described below in Sec. \ref{subsec:disk}.

The ratio $\lh/\lsoft$ (or equivalently the amplification factor, $A$, 
$l_{\rm h}/l_{\rm s} \equiv A - 1$) can be computed from the geometrical
parameters describing the plasma clouds:
$\mu_{\rm s}$ defined in eq.~\ref{equ:lsoftp}, the disk covering factor, $C$, and the
fraction of energy dissipated in the corona, $f$. 
We write
\be
A=\frac{L_{\rm soft}+L_{\rm diss}}{L_{\rm soft}},
\label{Adef}
\ee
where $L_{\rm diss}$ is the dissipation rate of gravitational energy in the corona
(the clouds), 
and $L_{\rm soft}$ is the luminosity of the disk radiation cooling 
the hot plasma.

We follow the calculations of B99, but we assume that only a
fraction of the available gravitational energy, $f\le 1$, is dissipated by the clouds
above the disk. The luminosity  of the hot clouds
is thus  
\be
L_{\rm diss}=f \epsilon \dot{M}c^2.
\ee
If the emitting plasma moves with high velocity, the luminosity transforms from
the source (plasma) frame to the observers frame  with $\delta^3$, where
$\delta=[\gamma(1-\beta\mu)]^{-1}$ is a Doppler factor (Rybicki \&
Lightman 1979), and $\mu=\cos\theta$, $\gamma^2=1/(1-\beta^2)$,
$\beta=v/c$ is the 
vertical velocity of the clouds (negative $\beta$ means movement toward the
disk). Also,  $\int_{-1}^{1}L_{\rm diss}(\mu)d\mu=L_{\rm diss}$, where
$L_{\rm diss}(\mu)$ is the angular distribution of the clouds luminosity in the
observer frame. Therefore, 
\be
L_{\rm diss}(\mu) = \frac{f{\epsilon}\dot{M}c^2}{2\gamma^4(1-\beta\mu)^3}.
\label{lmup}
\ee

The hard radiation emitted by clouds illuminates the accretion disk. A fraction $1-a$
($a$ being the disk albedo)
of it is thermalized in the disk, thus contributing to the disk
emission. We use $a = 0.2$ as appropriate for relatively cold matter 
(Haardt \& Maraschi 1993). As a
result the total disk luminosity is: 
\be
L_{\rm disk}=\epsilon\dot{M}c^2 (1-f)+(1-a)\int_{\rm -1}^{0}L_{\rm diss}(\mu)d\mu,
\label{equ:ldisk}
\ee
where the first term comes from the viscous
dissipation in the disk, and the second one represents the contribution
from the reprocessed radiation.

Not all the disk radiation cools the clouds, since only a fraction of the disk surface 
is covered by the clouds. We describe the fraction of the viscous
disk radiation intercepted by the clouds by introducing the disk covering parameter,
$C$. The fraction of the reprocessed radiation that returns to
the clouds is parametrized by the cloud geometry factor, $\mu_{\rm s}$, defined as in
B99 (see also Figure \ref{fig:geom}). The total
luminosity that provides cooling of the clouds may be now written as: 
\be
L_{\rm soft}=C(1-f)\epsilon\dot{M}c^2+(1-a)\int_{-1}^{-\mu_s}
L_{\rm diss}(\mu)d\mu.
\label{equ:lsoftp}
\ee
In such form, this equation allows for  only a small fraction of the disk being
covered by the clouds, opposite to what was done in Di Salvo et al.\
(2001), who assumed that the angular distribution of the viscous and
reprocessed disk radiation is similar, i.e.\ the clouds must cover most of
the disk's surface (cf.\ their equation (8)). In our approach, we expect $C \ll 1$
from the data modeling.

The amplification factor calculated in
the plasma comoving frame (Rybicki \& Lightman 1979; B99)
takes the following form:  
\begin{eqnarray}
A &=& 1 + f \left[C(1-f)\gamma^2\left (1-\frac{\beta(1+\mu_{\rm s})}{2}\right ) + \right. \nonumber \\
&+&\left.(1-a)\frac{f}{2\gamma^2}(1-\mu_s)\frac{1-\frac{\beta^2}{4}(1+\mu_s)^2}{(1+\beta)^2(1+\beta\mu_s)^2}\right]^{-1},
\label{Ap}
\end{eqnarray}
which for $\beta=0$ reduces to 
\be
A_{\beta \rm = 0} = 1 + \frac{f}{C(1-f) + 0.5 f (1-a)(1-\mu_{\rm s})}
\label{Ab0}
\ee

Given the input disk spectrum (described below, Section
\ref{subsec:disk}), the Comptonization procedure computes the
Comptonized component, and the component that escapes the clouds
without being scattered. 

\subsection{The disk component}
\label{subsec:disk}

The disk spectrum is assumed to be a disk blackbody with the radial
temperature profile of $T(r)=T_0\left(\frac{r}{r_{\rm
    in}}\right)^{-3/4}\left(1-\sqrt{\frac{r_{\rm
      in}}{r}}\right)^{1/4}$, 
where $r$ is in Schwarzschild units. The constant $T_0$ is derived
from the normalization condition that the total disk luminosity is
$L_{\rm disk}$. The disk photons that escape the system without
encountering the clouds on their way have luminosity $L_{\rm
  esc}=L_{\rm disk}-L_{\rm soft}$ (eq~\ref{equ:ldisk}, \ref{equ:lsoftp}).\\ 

The model is thus parametrized by:
\begin{itemize}
\item the mass of the black hole, $M$,
\item the accretion rate, $\dot{m} = \dot{M}/\dot{M_{\rm Edd}}$,
\item the fraction of gravitational energy released in the hot clouds, $f$,
\item the electron temperature, $kT_{\rm e}$,
\item the geometry of the clouds, $\mu_s$, 
\item the disk covering factor, $C$, and 
\item the vertical velocity of the clouds, $\beta$.
\end{itemize}

\section{Results}
\label{sec:results}
\indent
We compute the optical/UV/X-ray spectra based on the model described in the
previous section. From the spectra we compute spectral characteristics
such as the optical/UV luminosity, $l_{\rm UV} \equiv \log \left
( \nu L_{\nu} \right )$ at 2500 \AA\ in the rest frame, the X-ray
loudness, $\alpha_{\rm ox}$, the X-ray photon index, 
$\Gamma$, and compare them with observations. We
explore the model parameters space  to find theoretical parameters for which
modeled spectra match the observational data best. In the high temperature
case the X-ray spectrum deviates from the power-law-like shape due to
the first scattering effects, and we compute the photon index as the
index of the best power law fit to the model spectrum. 

Below, we discuss in details the results with respect to the data of
high-redshift RQQs. In particular we search for such values of the
theoretical model parameters which give $\Gamma < 2.3$ and $1.5
<\alpha_{\rm ox} < 1.8$.  Such a choice is motivated by the $3\sigma$
confidence intervals of $\Gamma$ given by V03
($1.7\leq\Gamma{\leq}2.3$), results of Vignali et al.\ (2003c) ($\Gamma 
= 1.86^{+0.41}_{-0.37}$), and B03 sample with the mean photon
index of $\Gamma = 1.50\pm 0.15$ (the error represents the 90\% confidence interval).
We choose the value of $\alpha_{\rm ox,min}
= 1.5$ for the lower limit motivated by the value within $3\sigma$
confidence interval found for RQQs in Bright Quasar Survey (see Brandt et
al.\ 2002b and references therein). The upper limit of 
$\alpha_{\rm ox,max} = 1.8$ was chosen based on the samples of B03 with 
$\alpha_{\rm ox}$=1.71$\pm$0.02 (the contributions to the error of $\alpha_{\rm ox}$
come only from errors of the X-ray photon index and the normalization of the 
power law fit to the data; also, the 2 keV flux in
the source frame was calculated from the 1 keV flux in the observer frame
assuming the photon indices found from fits, not the value of 2.2 as
assumed in B03), and V03 with $\alpha_{\rm ox}$=1.77$\pm$0.03 (we cite 
the value of $\alpha_{\rm ox}$ corrected by the authors
in a subsequent paper, Vignali et al.\ 2003c).

First, we focus on the case with $\beta
\equiv v/c = 0$, i.e. we assume that the clouds above the disk are
static. Next, we check how the mildly relativistic motion of the clouds
away from or towards to the accretion disk may affect the results.

\begin{figure*}
\epsscale{0.8}
\plotone{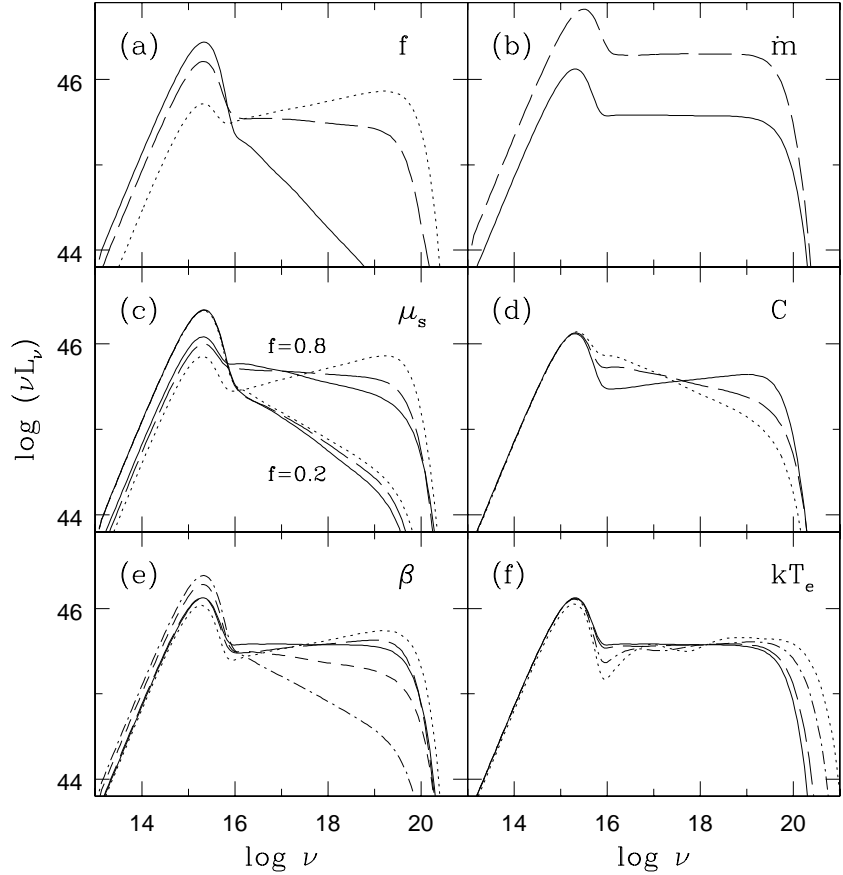}
\caption{Dependence of spectra on the model parameters. Unless stated
differently, the fixed parameters are: $M = 10^{10}\, M_{\odot}$,
$\dot{m} = 0.1$, $\mu_{\rm s} = 0.8$, $C = 0.2$, $f = 0.6$, $\beta =
0$, $kT_{\rm e} = 100$ keV; (a) the
  fraction of energy dissipated in the corona, $f = 0.1$ ({\it solid curve}), 0.5
  ({\it long-dashed curve}), 0.9 ({\it dotted curve}); (b) the accretion rate in Eddington units,
  $\dot{m} = 0.1$ ({\it solid curve}), 0.5 ({\it long-dashed curve}); (c) the cloud geometry
  parameter, $\mu_{\rm s} = 0.1$ ({\it solid curve}), 
0.5 ({\it long-dashed curve}), 0.9 ({\it dotted curve}); (d) the disk covering factor, $C = 0.1$
({\it solid curve}), 0.5 ({\it long-dashed curve}), 0.9 ({\it dotted curve}); (e) the vertical velocity of the
clouds, $\beta = -0.5$ ({\it dashed curve}), -0.2 ({\it dot-dashed
  curve}), 0 ({\it solid curve}), 0.2 ({\it long-dashed curve}), 0.5
  ({\it dotted curve}); (f) the electron temperature, $kT_{\rm e} =
100$ ({\it solid curve}), 150 ({\it long-dashed curve}), 300 ({\it
  dot-dashed curve}), 500 ({\it dotted curve}) keV.}
\label{fig:mgntpar}
\end{figure*}

Figure \ref{fig:mgntpar} shows the spectral changes due to the different model
parameters. 
The fraction of energy dissipated in the hot clouds affects both the
optical/UV luminosity and the X-ray slope (Fig.~\ref{fig:mgntpar}a).
Variations of the accretion rate have almost no influence on 
$\alpha_{\rm ox}$ and $\Gamma$ (Fig.~\ref{fig:mgntpar}b), but the optical/UV 
luminosity  is affected.  

Variations of the plasma temperature affect the high energy cut-off of
the X-ray continuum (Fig.~\ref{fig:mgntpar}f). In the spectra with
high temperature (here computed for $kT_{\rm e} = 300$ and 500 keV)
the characteristic feature of the first scattering can be seen in
model spectra. The models with low temperature (here $kT_{\rm e} = 100$ and
150 keV) have the optical depth of the order of
unity, while in case of high-temperature models the optical depth in
the plasma is significantly lower, which results in higher fraction of
the radiation that escapes the plasma without scattering. In the case
of significant coverage of the disk by the hot clouds this may result
in noticeable variations in the X-ray loudness. 

The rest frame ultraviolet
luminosity, $l_{\rm UV}$, depends only on the accretion rate, $\dot{m}$,
the fraction of energy  dissipated in the clouds, $f$, and the mass of the
black hole, $M$. For the chosen mass ($M=10^{10}\, M_{\odot}$), the
observed in high-$z$ RQQs $l_{\rm UV}$ is easily reproduced by any pair of $f$
and $\dot{m}$. Hence, we focus only on investigating the range of the X-ray
loudness, $\alpha_{\rm ox}$, and the X-ray slope, $\Gamma$, possible to
obtain from the model. 

The X-ray photon index does not depend on the accretion rate if all
remaining parameters are fixed. This can be seen in Figure
\ref{fig:mgntpar}b. Also the X-ray loudness (in the interesting range,
$\alpha_{\rm ox} \sim 1.5 - 1.8$) does not depend significantly on the
accretion rate for given cloud geometry, $\mu_{\rm s}$, fraction of
energy dissipated in the cloud, $f$, and the disk covering factor,
$C$. Hence, in this subsection we assume $\dot{M}$=0.1 $\dot{M}_{\rm Edd}$.  

For given $f$ and $C$ the X-ray spectra harden for higher cloud geometry parameter,
$\mu_{\rm s}$, i.e.\ for the cloud that is more compact in the sense of horizontal
dimension as seen from the disk. This effect is more pronounced for higher $f$ (see Figure
\ref{fig:mgntpar}c). More compact clouds intercept less of their own
reprocessed radiation. Less plasma cooling gives rise to hardening of
the spectrum. If $f$ is small, most of soft disk radiation comes from the
viscous dissipation in the disk, and the amount of the soft photons for
Comptonization in the cloud is determined by the disk covering parameter,
$C$. High $f$ means that in the soft disk flux the reprocessed hard cloud
radiation dominates. Therefore, any change of cloud geometry affects the
spectrum more in the high $f$ case than in the low $f$ case. 

\begin{figure*}
\epsscale{0.3}
\plotone{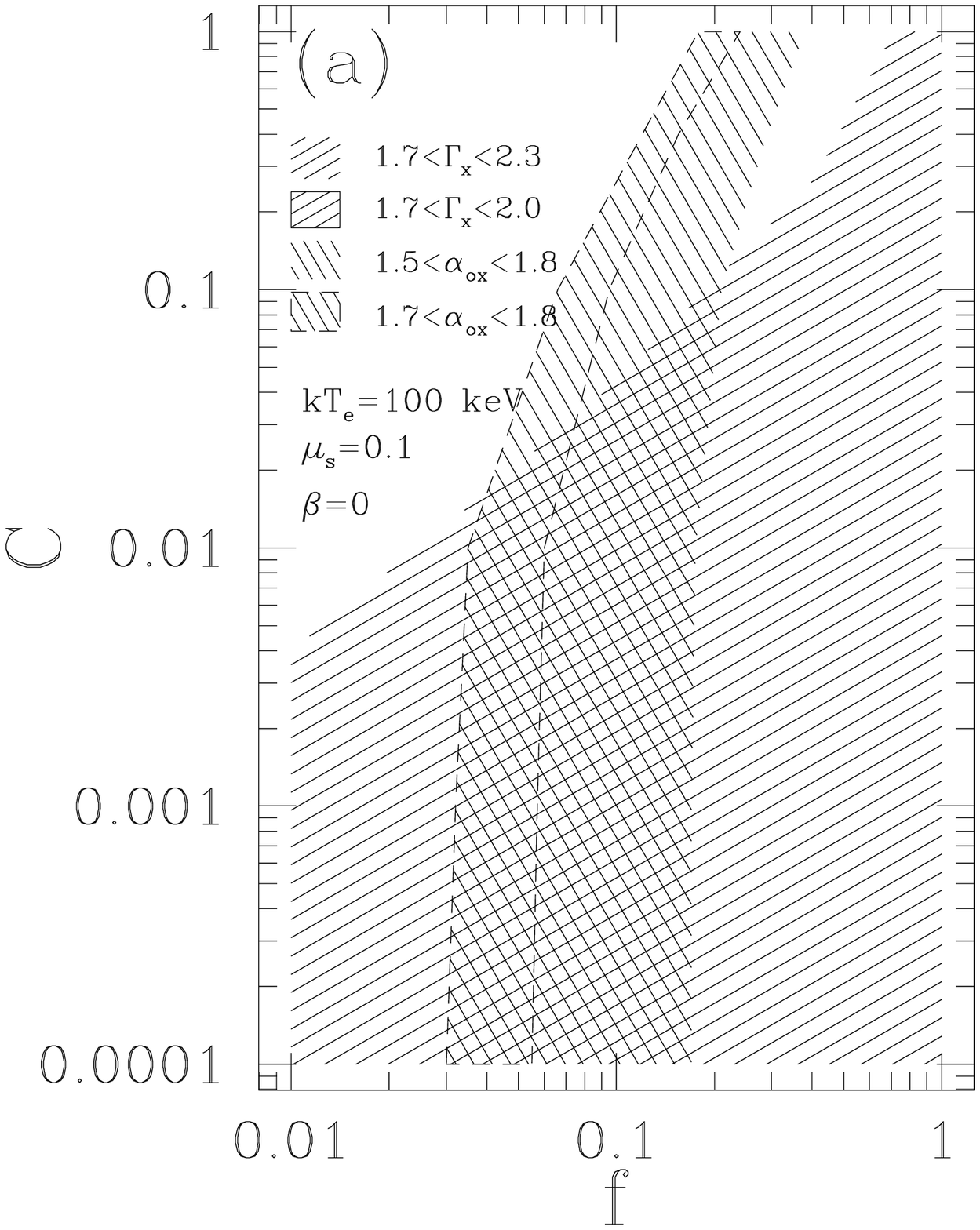}
\epsscale{0.3}
\plotone{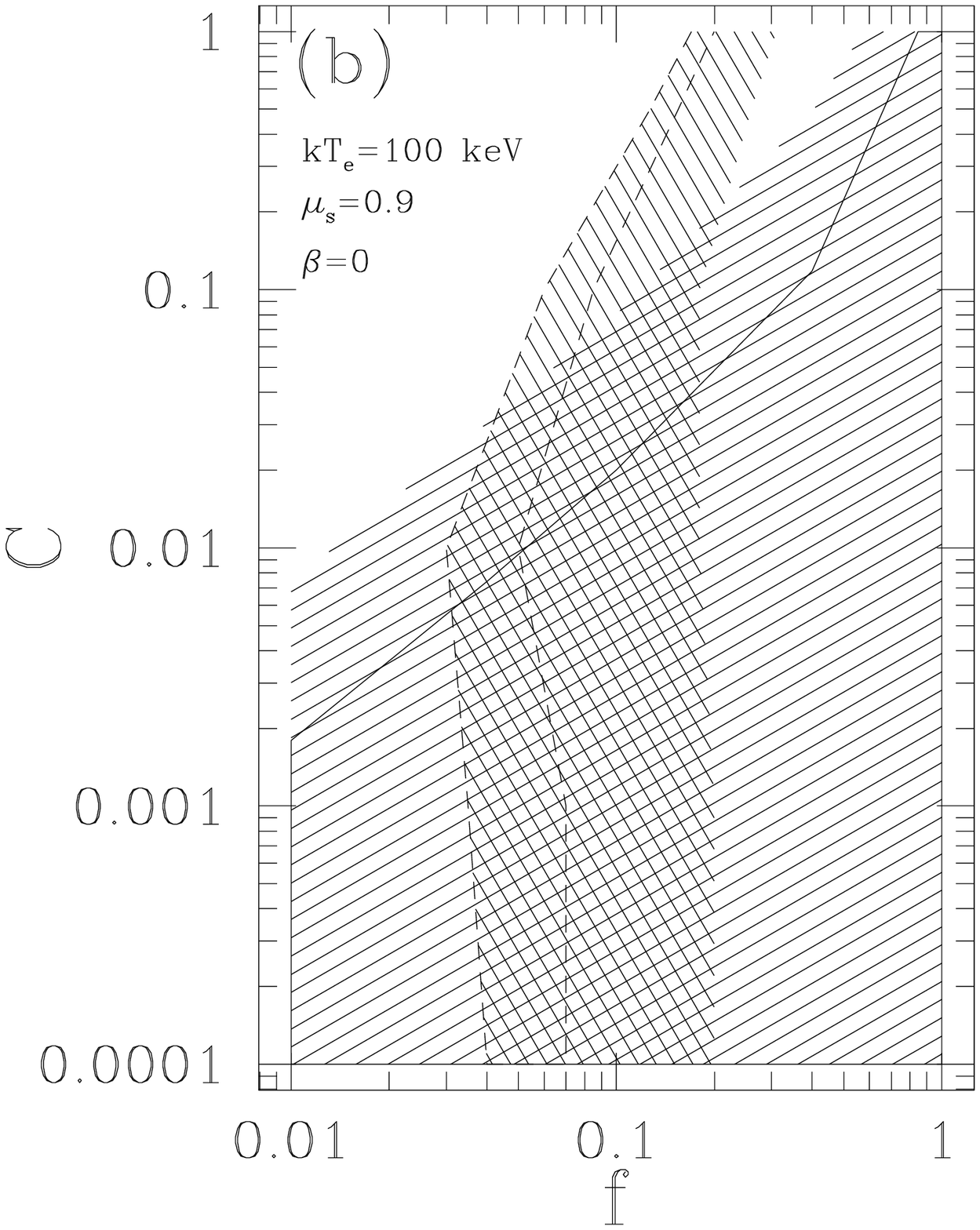}\\
\epsscale{0.3}
\plotone{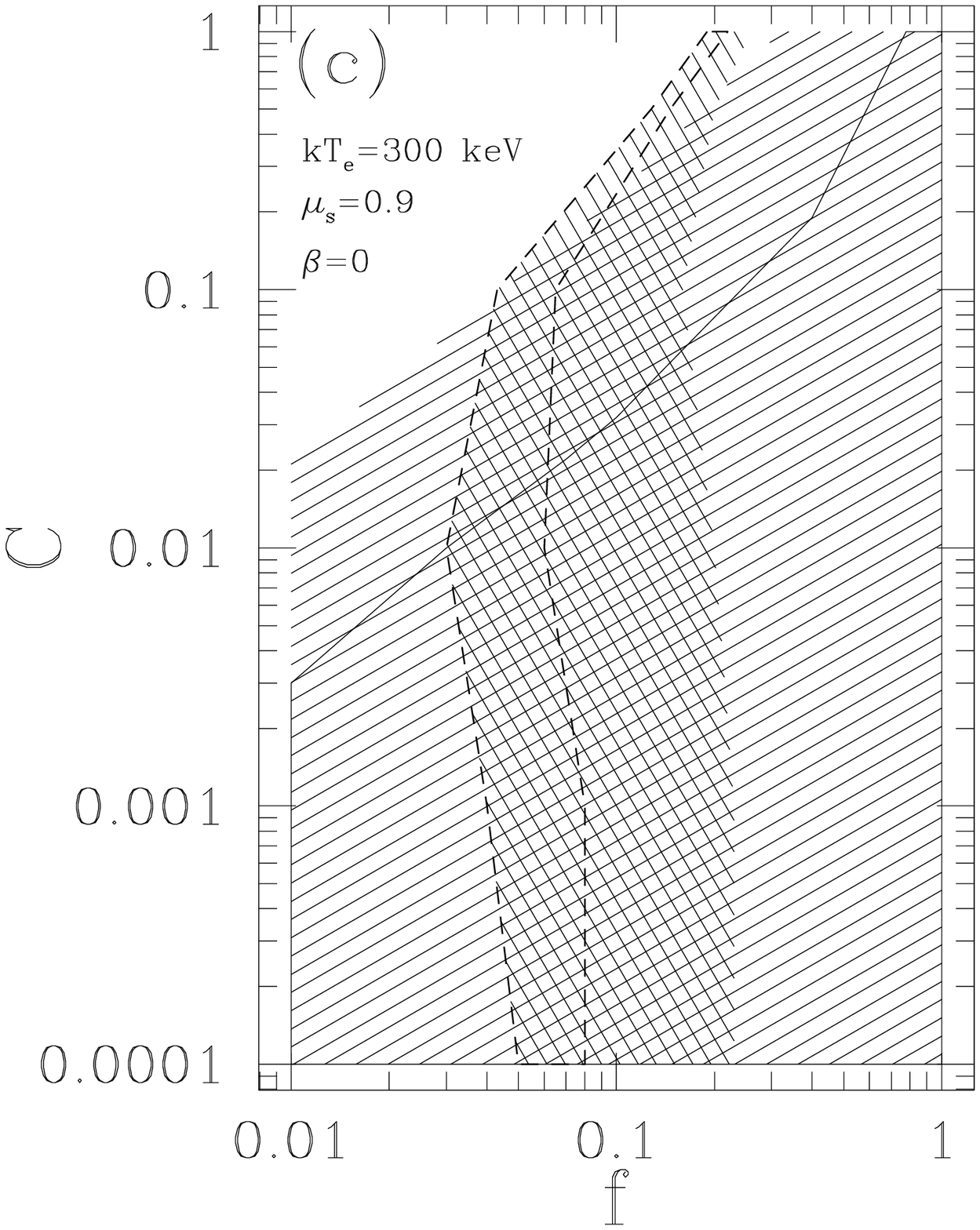}
\epsscale{0.3}
\plotone{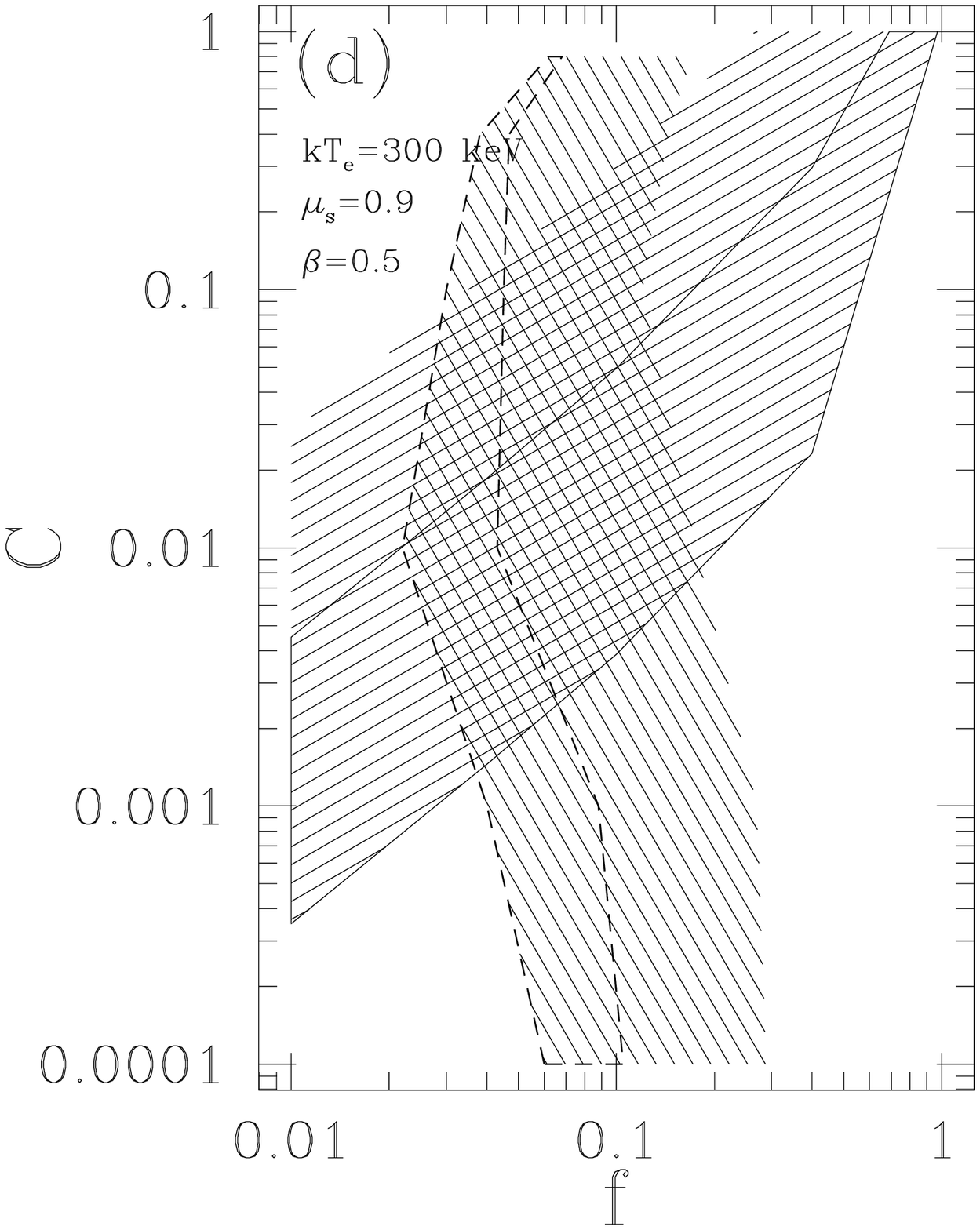}
\caption{Overlapping of regions with 
$1.5 < \alpha_{\rm ox} < 1.8$ and $1.7 < \Gamma < 2.3$ in the plane of 
the disk covering factor, $C$, and the fraction of energy dissipated in
the clouds, $f$. The computations were done for (a) $\mu_{\rm s}$=0.1,
$kT_{\rm e} = 100$ keV, $\beta = 0$, (b) $\mu_{\rm s} = 0.9$, $kT_{\rm 
e} = 100$ keV, $\beta = 0$, (c) $\mu_{\rm s} = 0.9$, $kT_{\rm e} = 300$
keV, $\beta = 0$, (d) $\mu_{\rm s} = 0.9$, $kT_{\rm e} = 300$ keV,
$\beta = 0.5$. The fixed parameters are $M = 10^{10}\, M_{\odot}$,
$\dot{M}=0.1\, \dot{M}_{\rm Edd}$. Extended clouds ($\mu_s = 0.1$)
produce spectra with $\Gamma \geq 2$ (a). In order to obtain spectra
with $\Gamma = 1.7$--2 the clouds should be compact (b,c) and/or move
away from the disk (d).} 
\label{fig:mgntreg}
\end{figure*}

In Figure \ref{fig:mgntreg} we present the main results of our
analysis. On the {\it disk covering factor - strength of the corona}
plane, we indicate regions with 
$1.5 < \alpha_{\rm ox} < 1.8$ and $1.7 < \Gamma < 2.3$. For
horizontally extended clouds
(with $\mu_{\rm s} = 0.1$) it is not possible to obtain spectra with the photon index
lower than approximately 2.2 (a).
Figure (b) shows that for vertically extended clouds the region with
the photon index of 1.7--2.0 appears in the parameter space.

Figure (c) illustrates the influence of the plasma temperature (the
optical depth) on the parameter space. Increasing the electron temperature does not
influence significantly the X-ray slope of the spectrum, which is determined
by amplification factor, whose value follows from global energy balance (see,
e.g., Beloborodov 1999ab). The X-ray loudness is mostly affected by the
temperature changes in the region of $C > 0.1$.

Spectra with X-ray loudness of $\alpha_{\rm ox} = 1.5$--1.8 are
produced if the hot clouds dissipate $\sim 3$--30\%  
of gravitational energy. The hardest and X-ray quietest spectra can be
produced for the fraction of energy released in the clouds of 3--8\%,
and the disk covering factor of $\leq 1$\%. 

A mildly relativistic outflow of the plasma causes hardening of
the spectra, whereas an inflow results in the softer X-ray photon index. This
is due to the relativistic beaming of the radiation. We illustrate this effect in
Figure \ref{fig:mgntpar}e. Figure \ref{fig:mgntreg}d 
shows the  transformation of the solutions for $\alpha_{\rm ox}$ and $\Gamma$
in the case of outflow with velocity of $\beta \equiv v/c =
0.5$. Spectra with the photon index lower than $\sim1.7$ can be
obtained only assuming the outflowing corona. However, the quality of
the data is not sufficient to draw the conclusions regarding the
plasma velocity since there is no information about the reflected
component. We discuss this problem further in the next Section.

\section{Application to the data}

\begin{figure*}
\epsscale{0.45}
\plotone{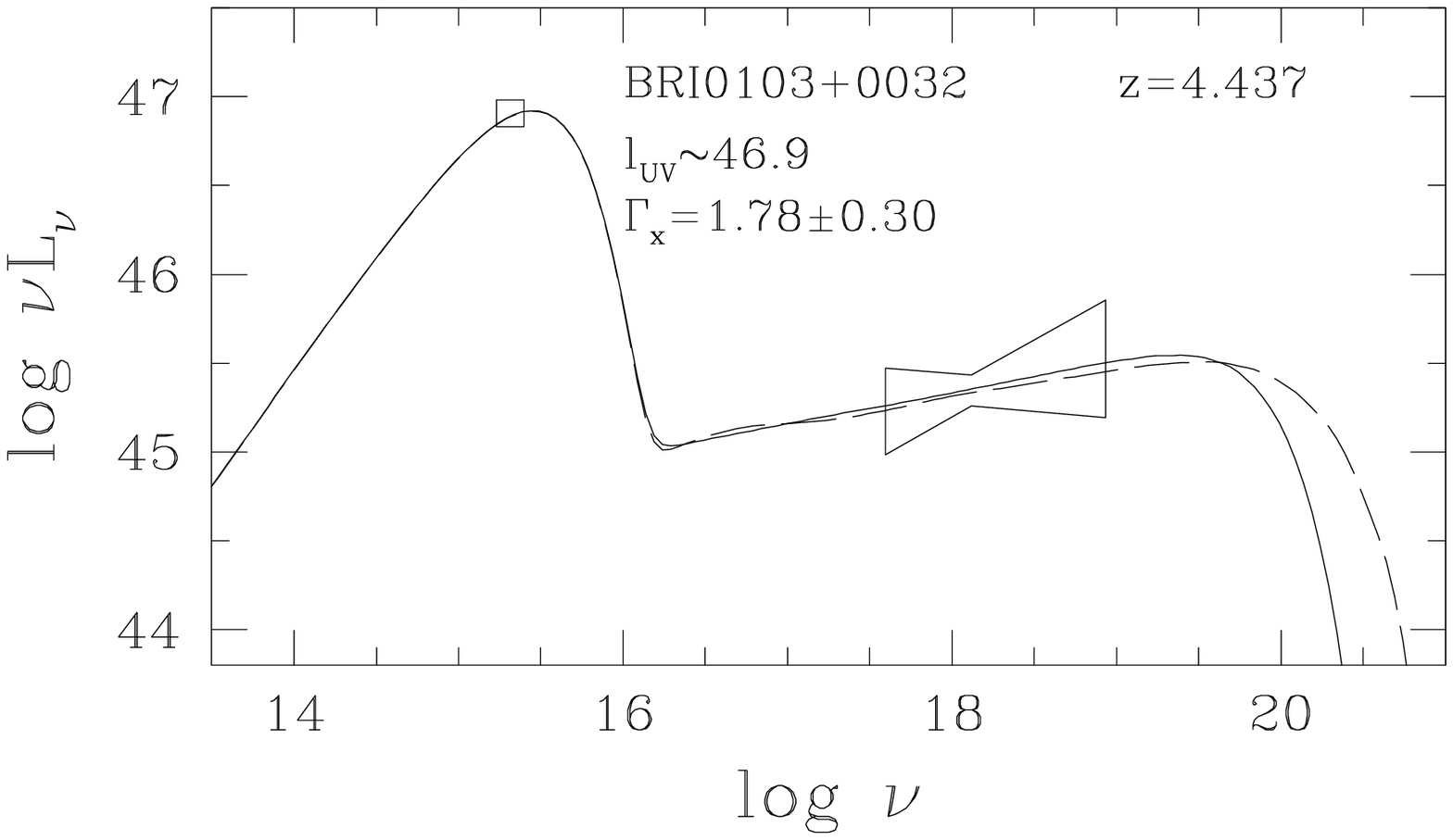}
\epsscale{0.45}
\plotone{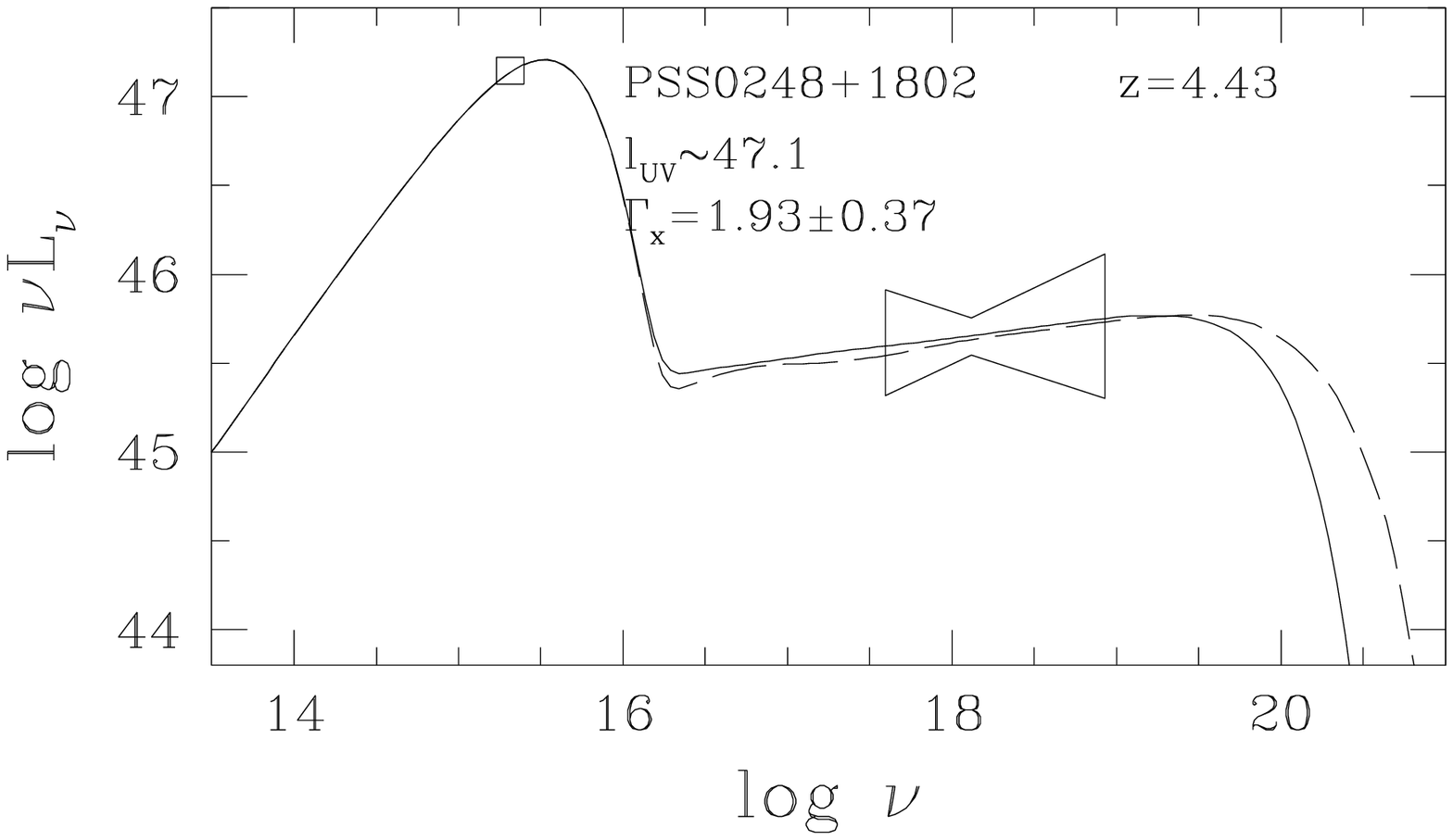}
\epsscale{0.45}
\plotone{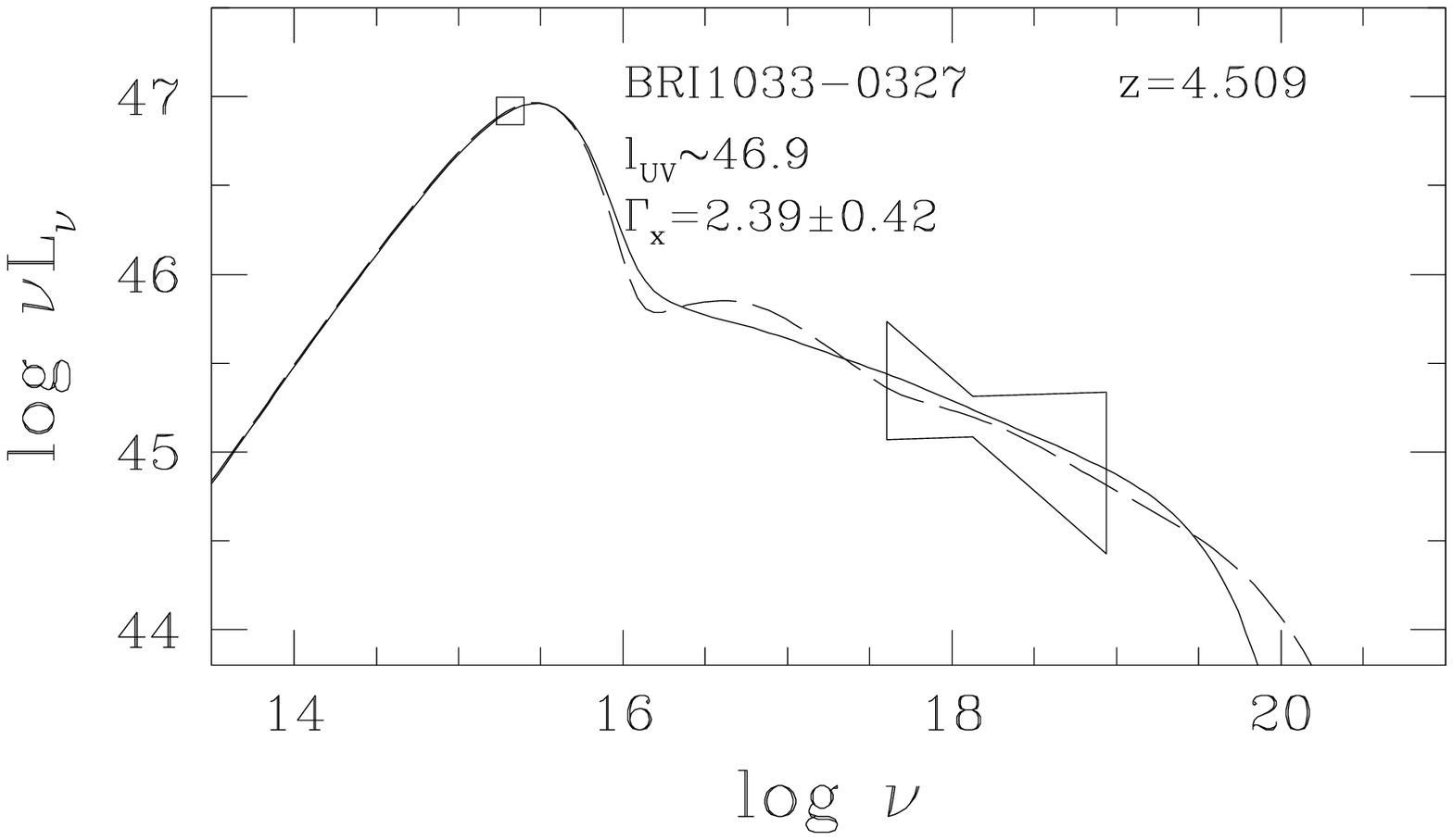}
\epsscale{0.45}
\plotone{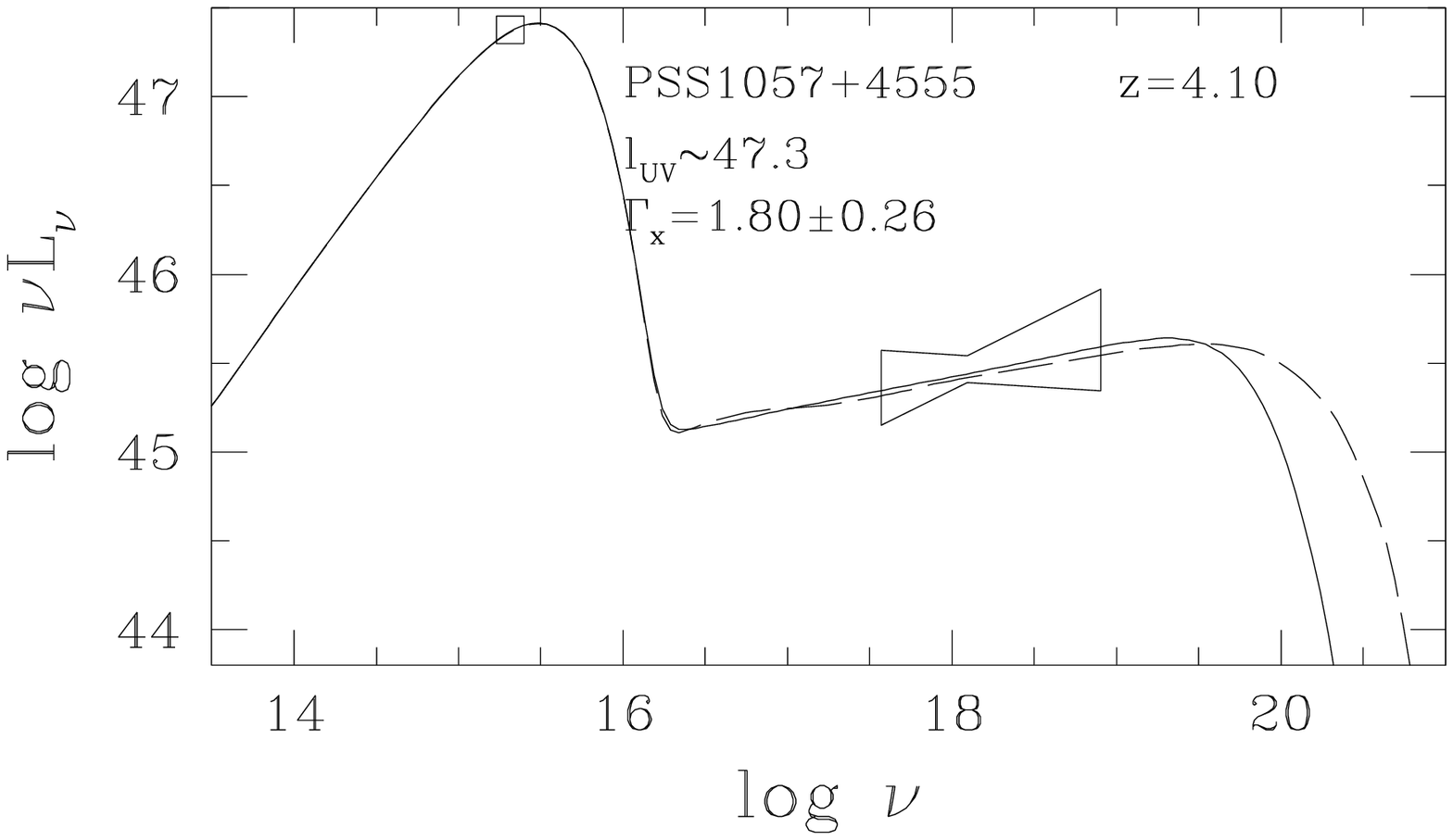}
\epsscale{0.45}
\plotone{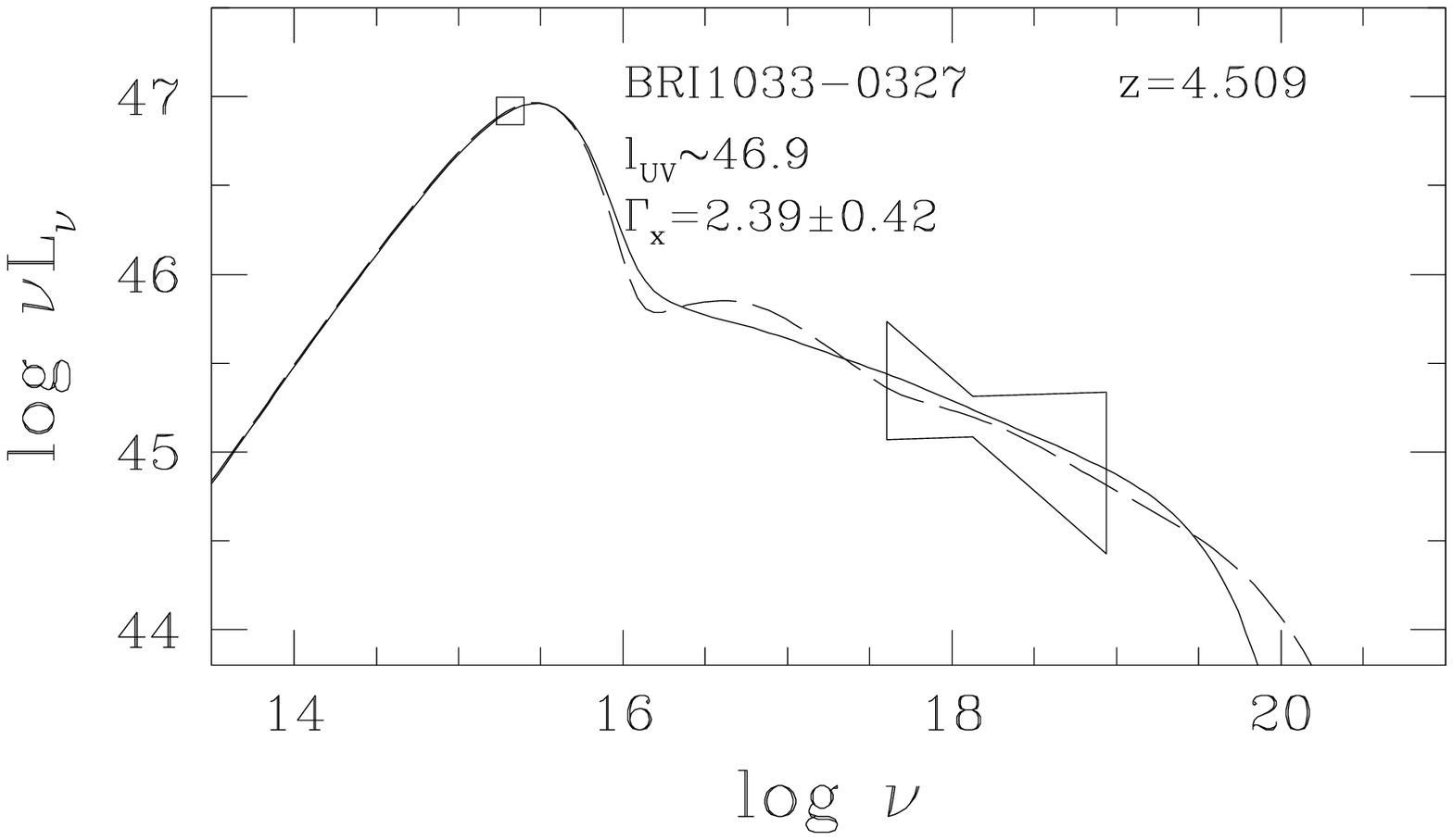}
\epsscale{0.45}
\plotone{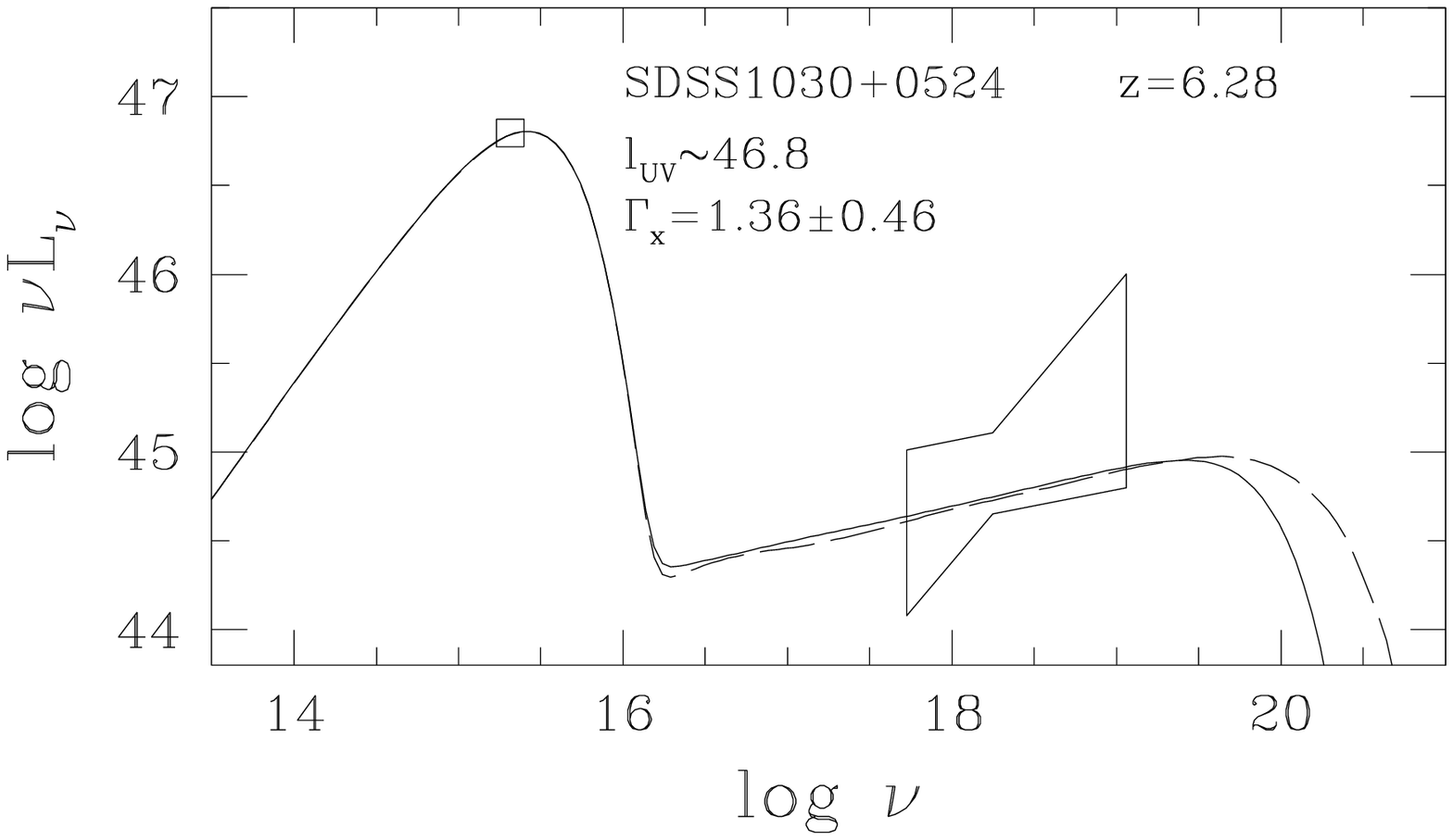}
\epsscale{0.45}
\plotone{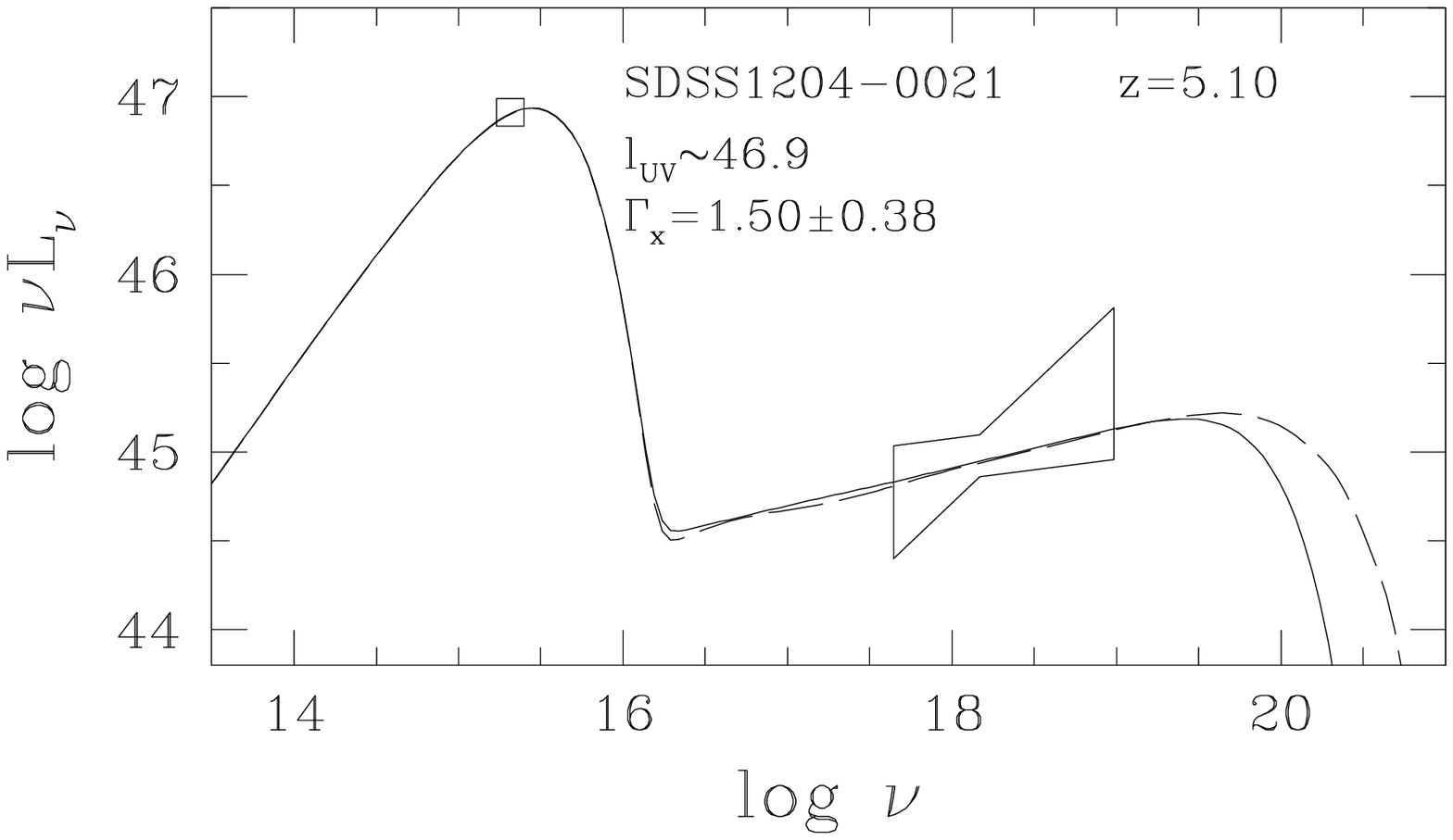}
\epsscale{0.45}
\plotone{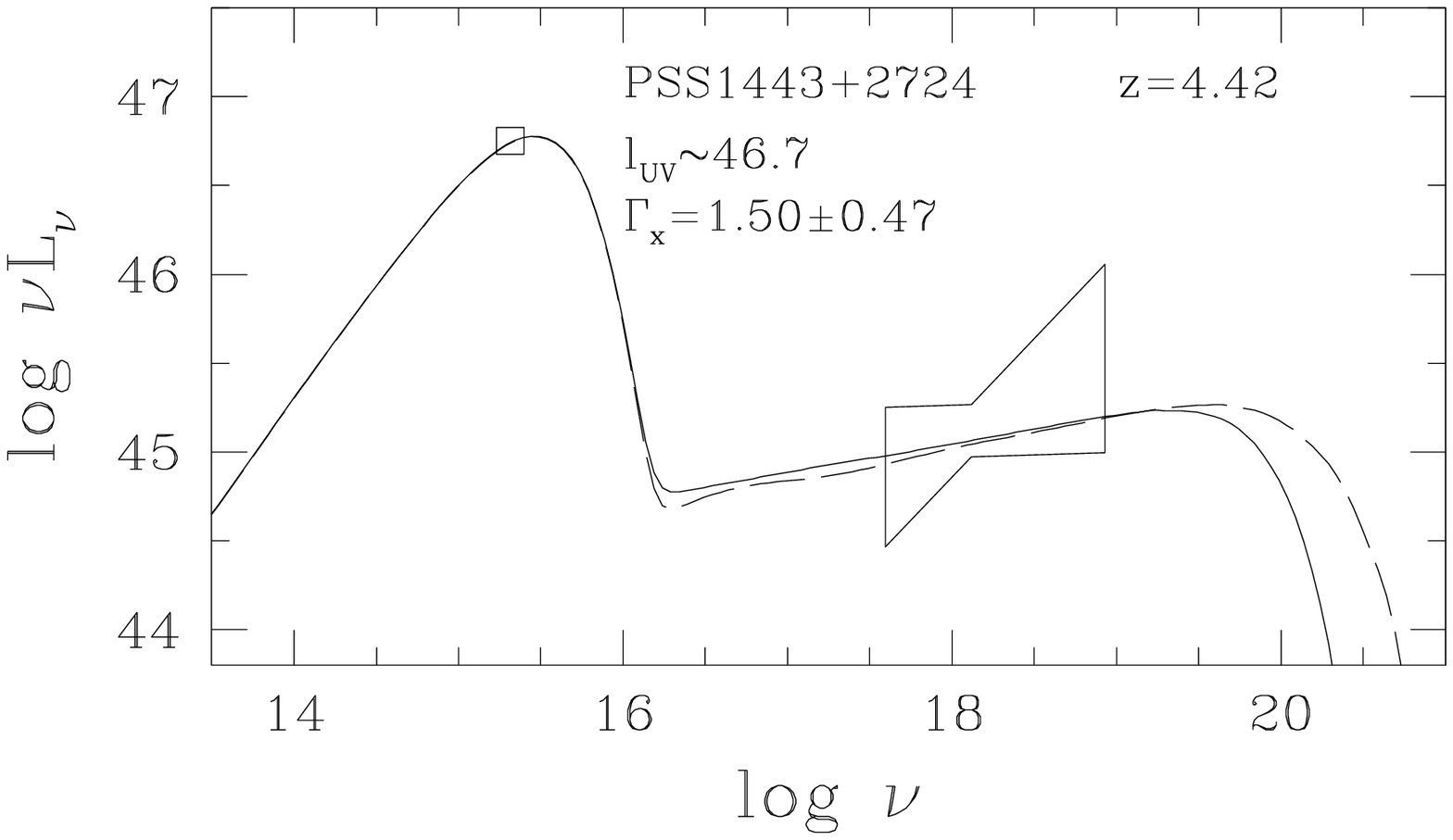}
\caption{Modeling the objects from the B03 sample in the static
  clouds case ($\beta = 0$) for $kT_{\rm e} = 100$ keV ({\it solid
  curve}), 300 keV ({\it dashed curve}). The parameters of the
  fits are listed in Table~\ref{tab:mgntb0}.} 
\label{fig:exmplsb0}
\end{figure*}

\begin{figure*}
\epsscale{0.45}
\plotone{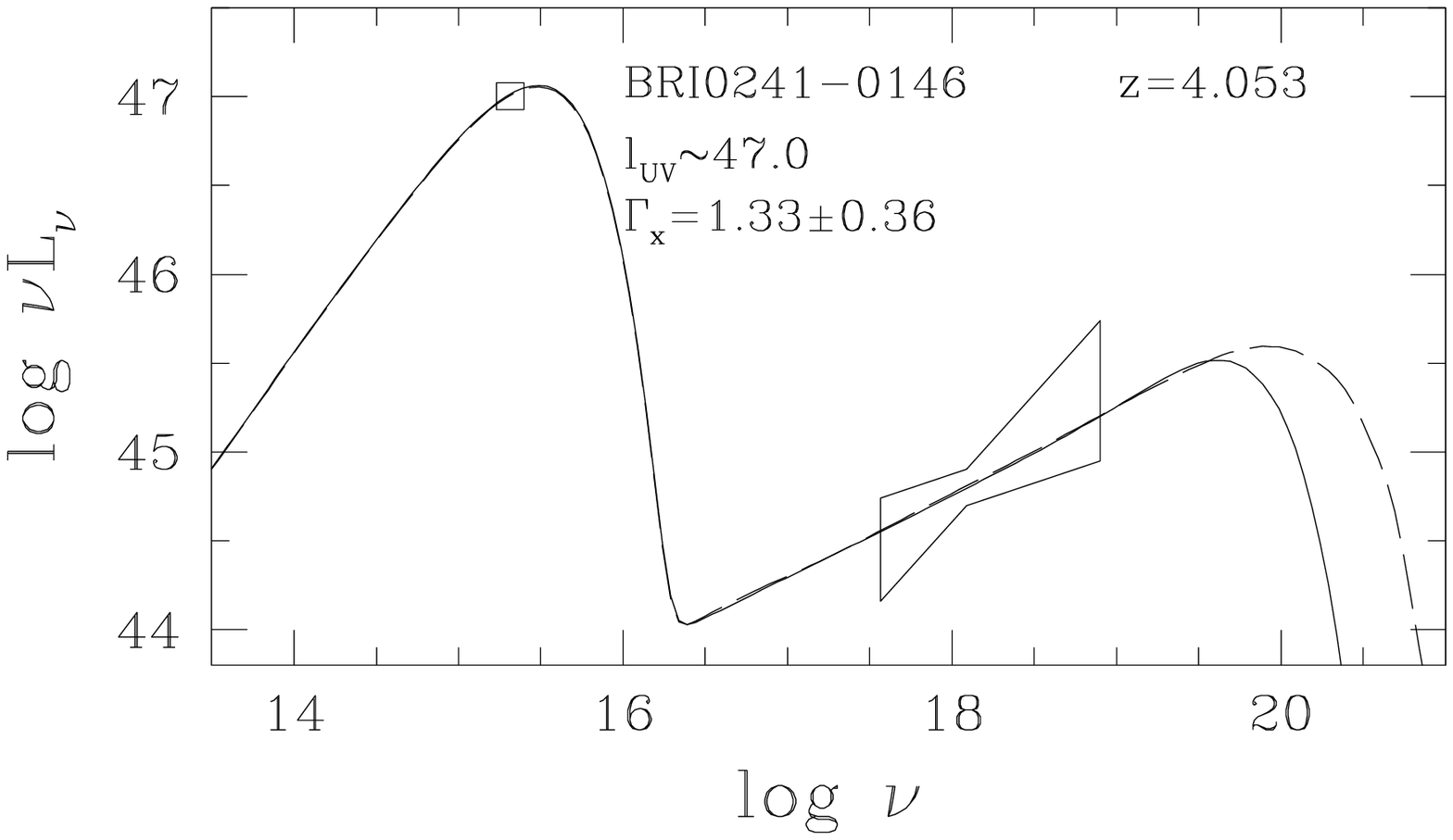}
\epsscale{0.45}
\plotone{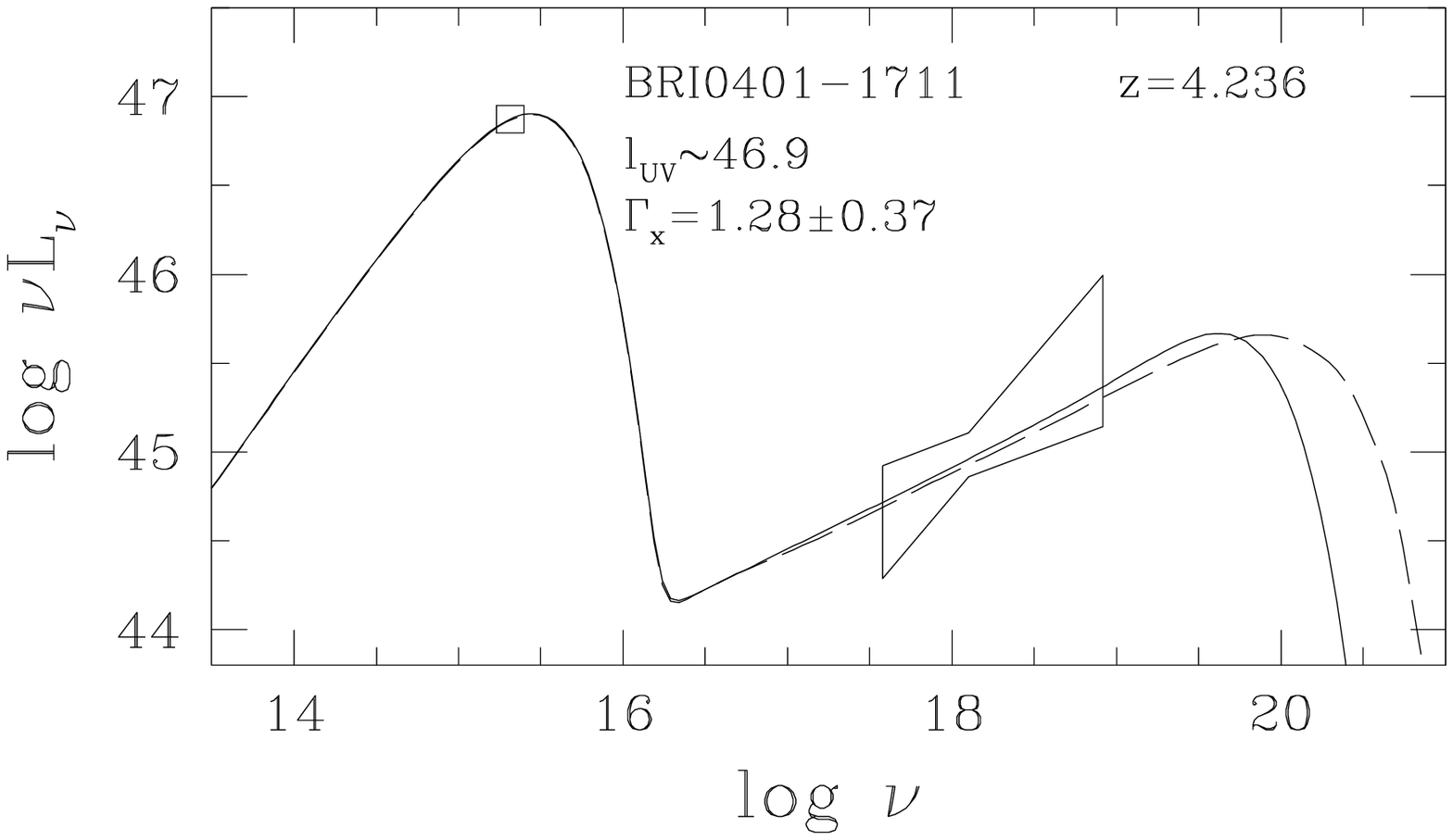}
\epsscale{0.45}
\plotone{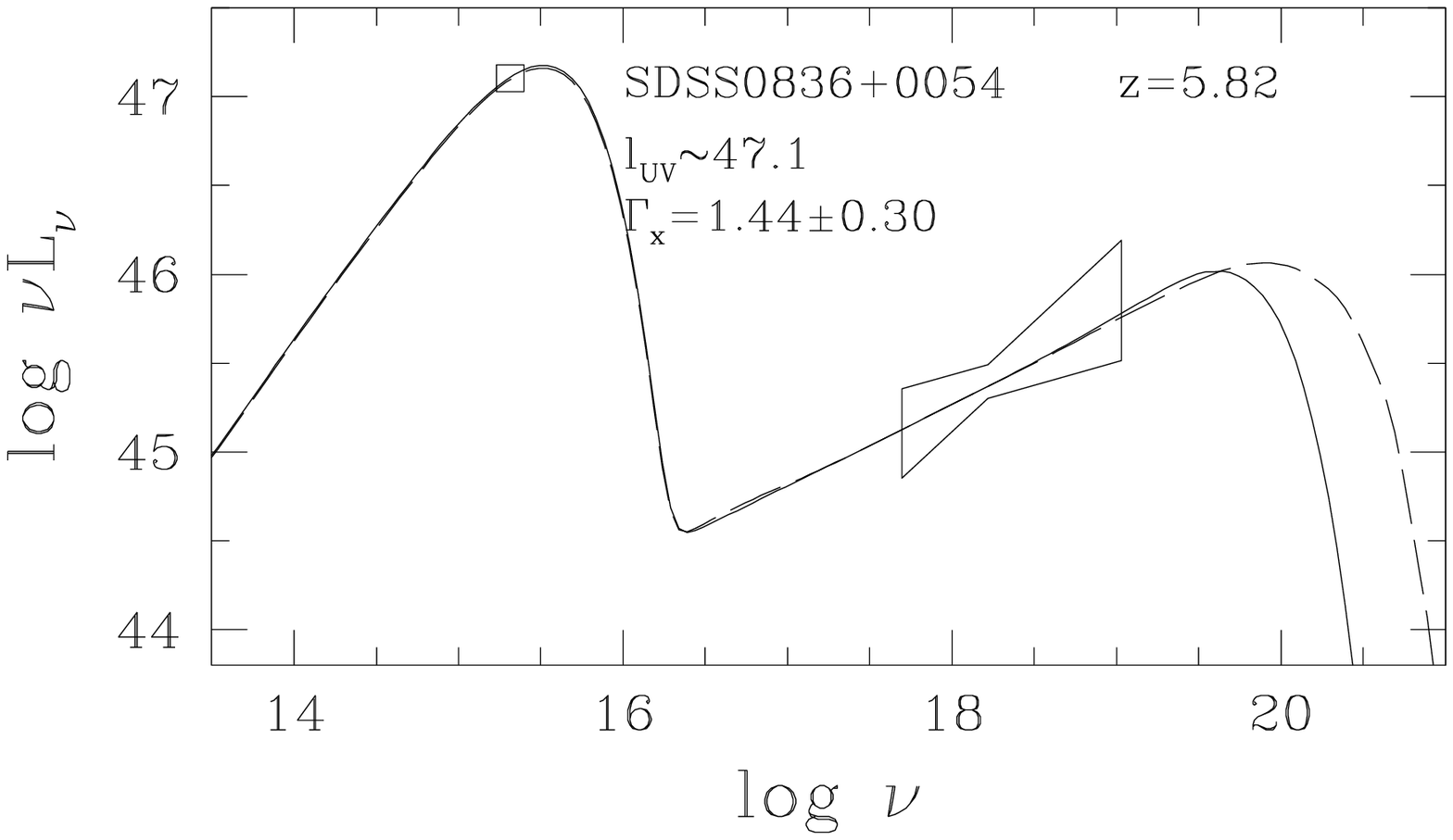}
\epsscale{0.45}
\plotone{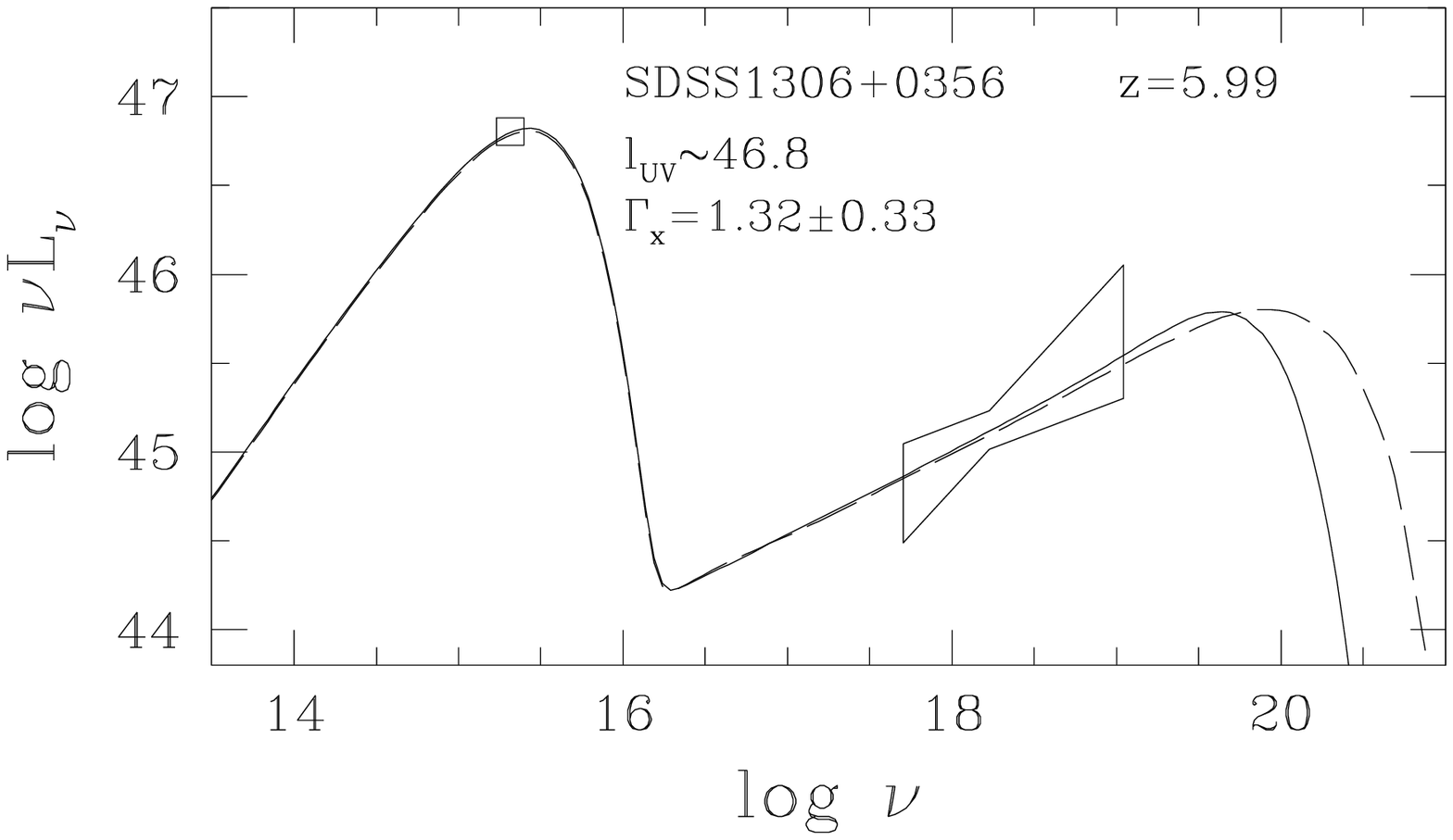}
\epsscale{0.45}
\plotone{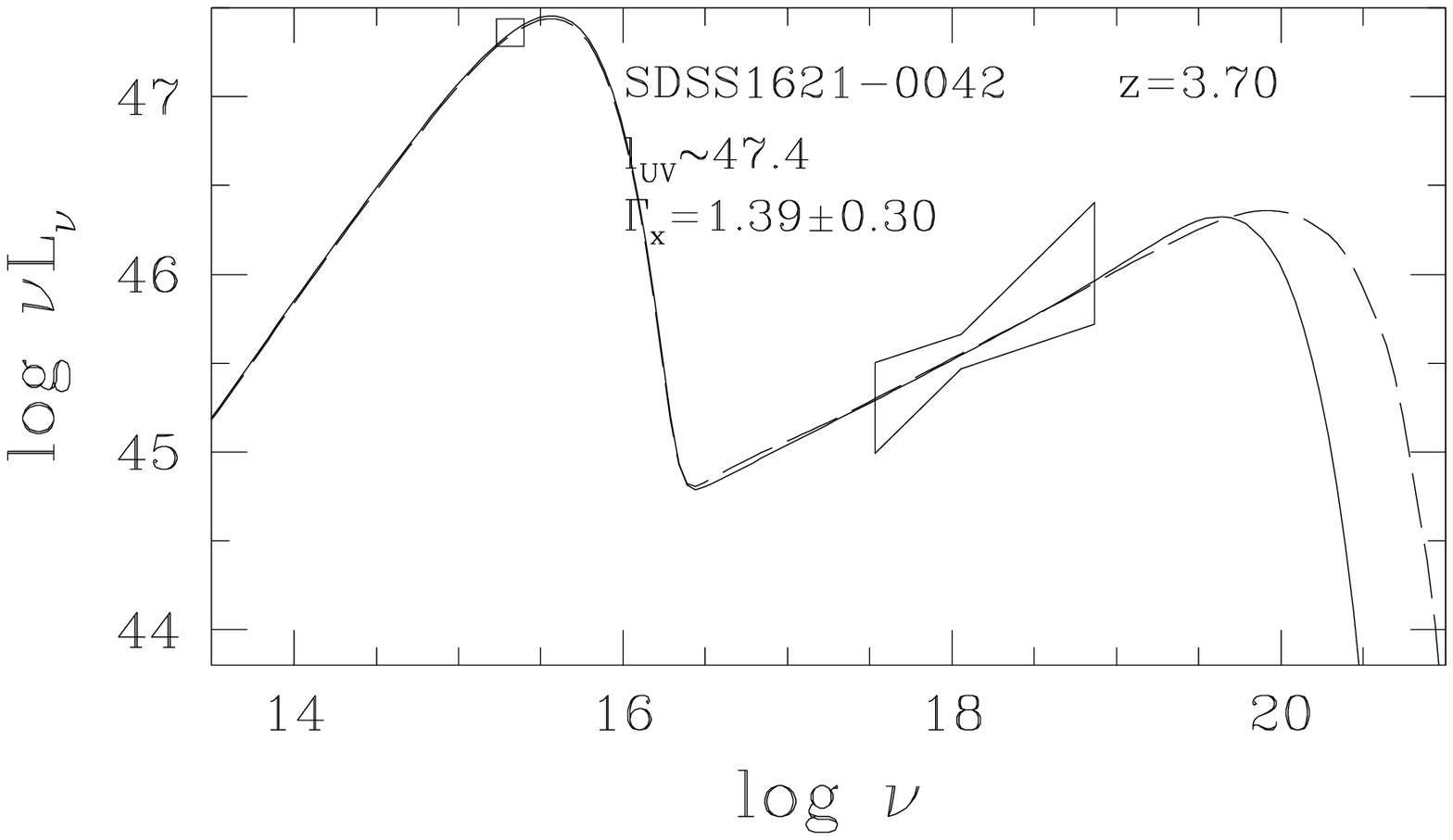}
\epsscale{0.45}
\plotone{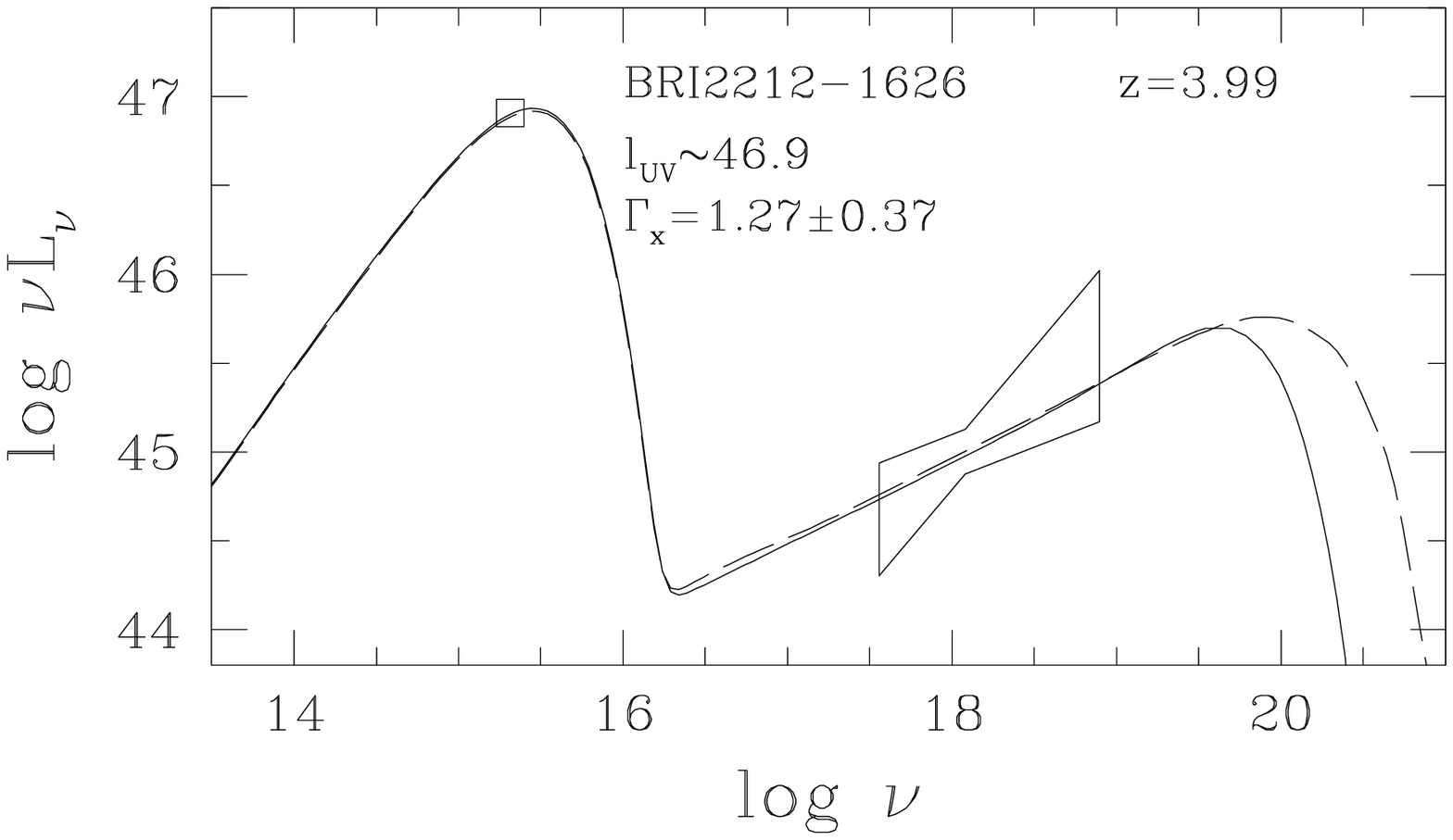}
\caption{Modeling the objects from the B03 sample in the outflowing
  corona case ($\beta = 0.5$) for $kT_{\rm e} = 100$ keV ({\it solid
  curve}) and 300 keV ({\it dashed curve}). The parameters of the
  fits are listed in Table~\ref{tab:mgntb05}.} 
\label{fig:exmplsb05}
\end{figure*}

We model spectra of all but two high redshift radio quiet quasars from the B03
sample. We omit two objects with the hardest spectra
characterized by the X-ray photon index as low as $\sim0.3$--0.7.
We do not perform a formal $\chi^2$ fitting to determine the best-fit
parameters and their confidence limits, but rather demonstrate the
parameter values which approximately reproduce the spectra.  With the
available data the results are not unique, and a number of solutions
is possible. The computed spectra are shown in Figures
\ref{fig:exmplsb0} and \ref{fig:exmplsb05}. The model parameters are
listed in Tables \ref{tab:mgntb0} and \ref{tab:mgntb05}. The objects
were observed by {\it Chandra 
X-ray Observatory} and the data were filtered to include the 0.3--6.5
keV energy range. The 1450\AA\ rest frame points 
were taken from the literature, as described in
details in B03. The ratio of the optical/UV and X-ray fluxes is
characterized by the X-ray loudness. In the case of high-$z$ quasars
the 2500 \AA\ flux used to calculate $\alpha_{\rm ox}$ must be
extrapolated from 1450
\AA\ flux known from observations, which involves a knowledge of the
optical/UV spectral index, $\alpha_{\rm UV}$ ($f_{\rm \nu} \sim
\nu^{\rm \alpha_{\rm UV}}$). To derive general properties of a
theoretical model or an observational sample, a mean value suggested
by observations can be used (as in Section \ref{sec:results}, or in B03
were $\alpha_{\rm UV} = -0.3$ was adopted (Kuhn et al.\ 2001)). However, it
does not apply to the spectral modeling of particular objects. Thus, in this
section we use the flux at 1450 \AA\ to normalize the spectrum, and
we list values of the spectral index between rest frame 1450 \AA\ and 2500 \AA,
$\alpha_{\rm UV}$, and the X-ray loudness, $\alpha_{\rm ox}$, computed
from the model spectra in Tables \ref{tab:mgntb0} and
\ref{tab:mgntb05}. The optical and X-ray 
observations were not simultaneous. However, Giveon et al.\ (1999) 
argue that the optical variability anti-correlates with the luminosity
of AGNs. Given that the high-$z$ RQQs are very luminous objects, one
would expect optical variability at a relatively small level.

Our modeling indicates that without information about the high energy
cut-off a grid of solutions exists for every object. For each object we
present fits to the spectra with $kT_{\rm e}$ of 100 and 300 keV. 

A method to distinguish between the low-$k T_{\rm e}$ high-$\tau$
and high-$k T_{\rm e}$ low-$\tau$ solutions could be obtaining high
signal-to-noise data of the Comptonized continuum at $\sim 0.1$--10 keV 
band (in the source frame). High-temperature low-$\tau$ Comptonized
continua show strong broad features from the first scattering, where
the shape of the seed photon spectrum is imprinted in the Comptonized 
spectrum (e.g.\ Stern et al.\ 1995). We have to keep in mind though
that the spectra are likely to be time variable and thus the observed
continua correspond to time averaged parameters, which may not contain
the specific features of a single temperature Comptonization.

From the modeling it follows that only less than 30\% of the energy is
dissipated in the corona.

The model is able to cover relatively wide range of $\alpha_{\rm UV}$
parameter from -0.1 to -0.4 (see Tables \ref{tab:mgntb0} and
\ref{tab:mgntb05}). Knowing exact 
value of this spectral index for particular objects would provide yet
another observable, which could help to determine the model parameters.

\subsection{Static hot plasma clouds.}

First we discuss the case with $\beta \equiv v/c = 0$, i.e.\ the hot clouds 
are static.
 
The data require a mass of the black hole of the order of $10^{10}
M_{\odot}$. The accretion rate is rather high, $\dot{M} > 0.2\,
\dot{M}_{\rm Edd}$. Certain degeneracy exists between the mass and the
accretion rate (i.e.\ one can explain the data with
either low masses and high accretion rates, or high masses and low
accretion rates); see BRI0103+0032 in Table \ref{tab:mgntb0}. In
addition, the geometrical parameters are also connected. For example, we
keep the strength of the corona constant and change the clouds
geometry parameter, $\mu_{\rm s}$, between 0.9 and 0.8. To model the
data in these two cases we need to adjust the disk covering factor,
$C$. Alternatively, one can keep the disk covering factor constant,
and adjust the strength of the corona. In this case, however, the
X-ray loudness fixed by the data additionally requires the plasma to have
lower optical depth.

\begin{figure*}
\epsscale{0.5}
\plotone{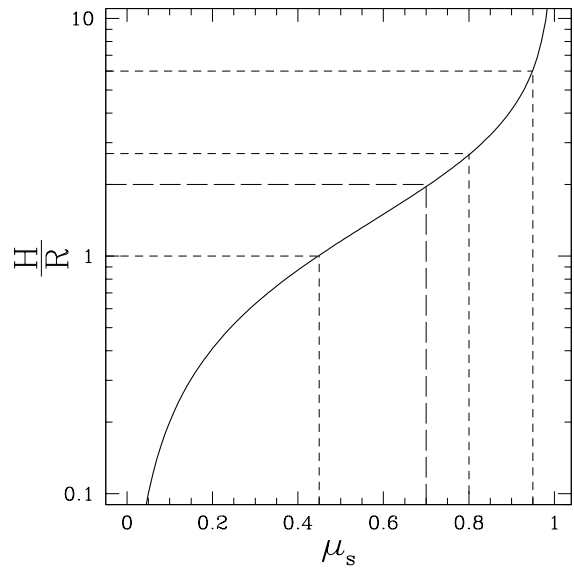}
\caption{Interpretation of the clouds geometry parameter, $\mu_{\rm
    s}$. $H$ denotes the heigth of the cloud, and $R$ denotes the cloud radius. The
clouds are approximately spherical for $H/R \sim 2$, which corresponds to
$\mu_s \sim 0.7$. For $\mu_s \sim 0.8-0.95$ the clouds are vertically
extended with $H/R \sim 3$--6. The higth and radius of the clouds are of
the same order for $\mu_{\rm s} \sim 0.45$. The plane-parallel geometry
corresponds to the case with $\mu_{\rm s} = 0$.}
\label{fig:mus}
\end{figure*}

To reproduce the observations the clouds need to be vertically
elongated with $\mu_{\rm s} > 0.4$, which to 
the first order corresponds to $H/R > 1$, where $H$ is the
height of the cloud and $R$ is its radius (see Figure
\ref{fig:mus}). The best fits presented in Figure~\ref{fig:exmplsb0}
and \ref{fig:exmplsb05} require $\mu_{\rm s} = 0.8 - 0.9$, i.e. $H/R
\sim 3 - 6$. For $\mu_{\rm s} < 0.4$ the spectra are too
soft. Increasing the strength of the corona, $f$, to harden the
spectra results in wrong relative normalization of the X-ray and
optical components.

The clouds intercept not more than $\sim 10$ \% of
the viscous disk radiation: the covering factor is typically of the
order of $10^{-4}-10^{-3}$. This means that the clouds cover rather
small fraction of the disk surface. Values of $C \ll 1$ are in 
agreement with our assumptions (see equation (\ref{equ:lsoftp})). We
perform also fits with $C = 0$ (which correspond to the case when the
soft photons for Comptonization come solely from the reprocessed
radiation) and the softest photon index alowed by
the data to provide a lower limit for $\mu_{\rm s}$ parameter.

In the case of the two objects with relatively soft
X-ray spectra ($\Gamma > 2.3$, BRI1033-0327, and PSS1317+3531) the geometry
of the clouds is very different. They can form horizontally extended clumps
with $\mu_{\rm s} < 0.5$. In addition, the configuration in which the
corona covers the whole surface of the disk like in the standard plane-parallel
geometry ($\mu_{\rm s} = 0$, $C = 1$) is marginally possible in the
low-temperature high-optical-depth case and can be realized in the
high-temperature low-optical-depth case.

\subsection{Moving hot plasma clouds}

Now we consider a possibility that the clouds
are moving with mildly relativistic velocities, outflowing for
  $\beta>0$ (e.g. BRI0103+0032 in Table \ref{tab:mgntb0}) and
moving toward the disk for $\beta<0$ (e.g. BRI1033-0327 in Table
\ref{tab:mgntb0}). However, the 
plasma velocity cannot be constrained by the present data because of
insufficient number of observables, which results in degeneracies
between the model parameters. It is not possible to uniquely determine the
outflow velocity, $\beta$, and the clouds geometry parameter,
$\mu_{\rm s}$, without information about the reflected component or
the high energy cut-off (B99; see also Di Salvo et al.\ 2001). The
amplitude of the reflected component does not depend on the clouds
geometry (except for a possibility of the reflected photons to be
destroyed while passing through the hot plasma). 
It is determined by the system inclination $\mu = \cos
\theta$ and the clouds velocity (B99). The reflection amplitude is
given by the following equation:
\be
R_{\rm refl} = \frac{(1+\beta/2)(1-\beta\mu)^3}{(1+\beta)^2}
\label{equ:reflamp}
\ee

This problem is illustrated by the fits to BRI0103+0032 (Table
\ref{tab:mgntb0}). We decrease the clouds geometry parameter, $\mu_{\rm
  s}$, from 0.9 to 0.5 keeping $C = 0.0035$ constant. The only way to compensate
for this change and obtain a spectrum with the same X-ray slope
($\Gamma = 1.88$) is to allow for an outflow with $\beta = 0.5$, and
adjust the strength of the corona, $f$, to reproduce the
correct relative normalization of X-rays and optical/UV emission. The
two fits (with $\beta = 0$ and $\beta = 0.5$) are equally good with
the present data. 

The objects in the sample with X-ray spectra characterized by the
photon index of $\Gamma \le 1.5$ can be modeled only assuming the
outflow (with the exception of the quasar at $z=6.28$ whose photon
index has large uncertainty). We present fits with $\beta = 0.5$
explaining their spectra (see Figure \ref{fig:exmplsb05} and Table
\ref{tab:mgntb05}).

\section{Discussion}

\indent
We have computed optical/UV/X-ray spectra from a two-component
accretion flow in geometry with the hot clouds of plasma (e.g.\ a
magnetically driven ``active corona'') distributed above an accretion
disk extending to the last stable orbit (Haardt et al.\ 1994; 
Stern et al.\ 1995). We assumed that the soft disk
blackbody-like radiation was Comptonized in the hot plasma clouds, and
the electron energy distribution in the hot plasma was thermal. We
investigated a parameter space where the models can reproduce the
observed spectra of high redshift radio quiet quasars, and
 we applied the models to the data of RQQs from the sample compiled by B03.
The observed spectra can
be characterized by three observables: (1) the luminosity at 2500 \AA\
in the rest frame of the source, $l_{\rm
UV}\equiv{\log}{\nu}L_{\nu}\sim 46$--47, (2) the X-ray loudness,
$\alpha_{\rm ox}\sim 1.5$--1.8, (3) the X-ray photon index,
$\Gamma\sim 1.7$--2.3. These parameters imply that, for most of objects
from the sample, the spectra are dominated by the disk thermal component,
but the hard X-ray continua are rather hard.

Modeling the spectra
we found that the angular size of the Comptonizing clouds as seen from
the accretion disk should be small, $\mu_{\rm s} \equiv \cos \theta
\sim 0.8$--0.9 for best fit models, which corresponds to the ratio of
height to radius $H/R \sim 3$--6 (Fig.~\ref{fig:geom}). 
The clouds dissipate up to $\sim 30$\%
of gravitational energy, and intercept less than $\sim 10\%$ of viscous disk
radiation.

In Paper I (the case of hot inner flow geometry) we have found 
that the plasma seems to have rather low optical depth (high electron
temperature). Here we also find this kind of solution (with $kT_{\rm
  e}$ of 300 keV and $\tau \sim 0.02$--0.5) possible. However, the
modeling is not unique because of to small number of observables.
For each object a grid of models exists parameterized by, for example,
plasma temperature. Solutions with low plasma temperatures
and optical depths $\sim 1$ are possible, and these parameters are 
comparable to those of Seyfert galaxies and
XRBs (although the broad band spectra are rather different, see below). 
These solutions however require very low energy dissipation in the hot
clouds and small disk covering factor.

Possible bulk outflow velocity of the plasma (B99) is not constrained
by the present data of high-$z$ RQQs because of the lack of
information about the reflected component. Malzac et al.\ (2001)
discuss effects of scattering and attenuation of the reflected
component traveling through the active region in patchy corona
geometry, which may affect the equation (\ref{equ:reflamp}). They
assumed cylindrical shape of active coronal regions, which results in
different numerical coefficients in $\Gamma(A)$ dependence
(see Beloborodow 1999b) than in the case of spherical clouds, but the overall
trends should be similar. They concluded that these effects are strong
for small $H/R$ ratios (which corresponds to small $\mu_s$) reducing
the amount of reflection, and negligible for high $H/R$ ratios (high
$\mu_s$). In the static case they obtained hard spectra and $R_{\rm
  refl}=1$ for 
high $H/R$, and soft spectra and $R_{\rm refl}<1$ for small $H/R$. Thus, the
static patchy corona produces $R_{\rm refl}-\Gamma$ anticorrelation instead
of the observed correlation for X-ray Binaries and Seyferts (Zdziarski
et al.\ 1999). The correlation can be obtained if we consider 
a possibility of the clouds outflow/inflow in the model. 
There exist no data for high-$z$ RQQs to
check whether the $R_{\rm refl}-\Gamma$ correlation holds also for these objects.

The two models considered in this paper and Paper I give similar
estimates on the mass of the black hole in high-$z$ RQQs. The required
mass is of the order of (0.5--1.5)$\times 10^{10} M_{\odot}$. The
predictions on the fraction of energy dissipated in the hot plasma
above the disk are slightly different. The patchy corona model allows for
as much as $\sim$30\% of the dissipation to take place in the hot
plasma, whereas in the model with truncated disk and
hot inner flow as described in Paper I only about 10\% of the energy
may be dissipated in the corona above the disk (unless the spectra are soft, with
$\Gamma > 2.3$). The accretion rates are comparable in both models,
$\dot{M} > 0.20\, \dot{M}_{\rm Edd}$.

Spectra of objects in our sample are rather different from spectra
of local Seyfert galaxies. The quasars spectra are strongly disk-dominated,
yet the X-ray continua can be extremely hard. There is also greater
variety of hard X-ray spectral slopes, despite relatively narrow 
range of mass accretion rate. In local Seyfert 1 galaxies the disk
component appears to carry comparable  luminosity as the hard X-ray
component (e.g.\ NGC 5548, Chiang \& Blaes 2003; MCG-6-30-15, Reynolds 
et al.\ 1997), while the X-ray slope is
in the range 1.7--2. On the other hand, spectra of Narrow Line Seyfert 1 
galaxies, thought to 
accrete at rather high $\dot m$, are disk dominated (e.g.\ Puchnarewicz 
et al.\ 2001), but their X-ray continua are rather soft, $\Gamma>2$
(Brandt et al.\ 1997; Leighly 1999; Janiuk, Czerny \& Madejski 2001).

Even more interesting is the comparison
of the distant quasars with black hole binary systems in their 
high $\dot m$ spectral states (High State and Very High State, see
review by Done 2002). In these states the ratio of fluxes in the hard 
Comptonized and disk components can be anywhere between $\approx 0$ and 
$\approx 1$, but the hard X-ray slope is $\Gamma \ge 2$ 
(e.g.\ Done \& Gierli\'{n}ski 2003). 
Such strongly disk-dominated spectra with very hard Comptonized continua
are rarely observed. This has prompted Gierli\'{n}ski \& Done (2004)
to propose that the spectra of quasars are strongly affected by 
photo-absorption. The intrinsic spectra would be similar to black hole 
binaries in (Very) High State, but strong absorption, possibly from
a wind, results in apparent hard continua, with an apparent strong
soft X-ray excesses. Obviously, if their idea is correct, the
parameters of an active corona would be completely different from what we
have determined. In particular, the high energy X-ray continuum in XRB
 is usually explained as nonthermal Comptonization, since it is observed
up to $\sim 1$ MeV without any significant spectral breaks (Grove et al.\ 
1998).

Vignali et al.\ (2003b) report on the observed anti-correlation between the
X-ray loudness (defined in their paper as negative) and the optical/UV
luminosity in high-$z$ RQQs. In Paper I and in the present paper we defined
the X-ray loudness as positive. Below we discuss qualitatively, how
the correlation can be obtained within the framework of the two
geometries, the hot inner flow (Paper I) and the patchy corona
(this paper). 

In the model studied in Paper I (the
truncated disk with the hot inner flow), changes of the disk
truncation radius provide a natural qualitative explanation for this
correlation. If the disk truncation radius becomes smaller (i.e.\ the
inner edge of the disk moves towards the 
last stable orbit) the optical/UV luminosity rises. Also, the X-ray
spectrum becomes softer which causes a drop in the 2 keV rest frame
flux. Both higher optical/UV luminosity and lower 2 keV rest frame
flux cause a rise in the X-ray loudness. As a result a correlation
between optical/UV and the X-ray loudness can be obtained in the
model. The other model parameters that influence the optical/UV
luminosity are the accretion rate and the strength of the
corona. However, in the hot inner flow model the corona above the disk
is required to be weak with $f \leq 10$\%, and the accretion rate only
slightly influences the X-ray loudness, so their impact on the
observed correlation less significant than the change in the
truncation radius.

In the patchy corona geometry (this paper) variations in the
strength of the corona formed by hot clouds can account for the
correlation between the X-ray loudness and optical/UV luminosity. If
smaller fraction of gravitational energy is dissipated in the clouds,
the disk thermal radiation is stronger and the optical/UV luminosity
rises. At the same time the X-ray spectrum becomes softer. Thus 
the $\alpha_{\rm ox} - l_{\rm UV}$ correlation can be
obtained. Fluctuations of the clouds geometry parameter, $\mu_{\rm s}$
and the accretion rate, $\dot{m}$, also affect the optical/UV
luminosity. However, based on our modeling of the quasars' data 
a narrow range of $\mu_{\rm s}$ (0.8--0.9) is required, and in
addition the X-ray loudness does not vary significantly with
$\dot{m}$. Thus the energy dissipation in the corona could be the main
driver for $\alpha_{\rm ox} - l_{\rm UV}$ correlation in this model.

We note that the observed value of the X-ray loudness depends on the
assumptions concerning the optical/UV spectral index and the X-ray
slope.  The optical/UV spectra are affected by the intrinsic reddening
which is usually unknown, so the UV spectral index has large
uncertainties. Kuhn et al (2001) provides the best to date high
redshift quasars optical/UV spectra and their value of $\alpha_{\rm
UV}$=-0.3 is different from the mean value of the Sloan Digital
Sky Survey quasars sample of -0.79$\pm0.34$ (Fan et al 2001).
Brandt et al.\ (2002a) analyzed X-ray spectra of the
three highest redshift quasars known ($z = 6.28, 5.99, 5.82$) assuming
the average value for the photon index derived by V03 ($\Gamma \sim
2$) and $\alpha_{\rm UV} = -0.5$ (see references in Brandt et al.\
2002a). They obtained $\alpha_{\rm ox}
\sim 1.68, 1.60, 1.58$, respectively. B03 used $\Gamma = 2.2$ (Laor et
al.\ 1997) and $\alpha_{\rm UV} = -0.3$ (Kuhn et al.\ 2001) for the
same objects and obtained $\alpha_{\rm ox} \sim 1.61, 1.54, 1.55$,
respectively. We used the photon indices provided by B03 based on the
spectral fit to the {\it Chandra\/} data. Our spectral model applied
to these quasars gave spectra with the X-ray loudness of 1.78--1.79,
1.69, and 1.66--1.70, respectively (Tables \ref{tab:mgntb0} and
\ref{tab:mgntb05}). As shown in this paper and in Paper I, the
$\alpha_{\rm ox}$ parameter is one of the most important observables
for high-$z$ RQQs and it is important to understand its dependence on
$\Gamma$ and $\alpha_{\rm  UV}$. 

Obviously, tighter constraints on model parameters could be obtained
with larger number of observables. The most promising at the present
is the usage of optical/UV data for better modeling the optical/UV
spectral index. This would allow to constrain better the mass and the
accretion rate in the models. Another diagnostic would be measurements
of the amount of reflection which could discriminate between uniform
and non-uniform corona, or truncated accretion disk scenario. In
addition, the observations of the amount of broadening of the iron
K$\alpha$ line can provide estimates on the truncation
radius. The observation of the high energy cut-off in the hard X-ray
data would dramatically improve our understanding on the high-$z$ RQQs
spectra. However, the two latter tests require good quality data at
least up to 30 keV in the rest frame, which might be possible with the
future missions such as Constellation-X or NuSTAR for example.

\section{Conclusions}
\label{sec:concl}

In this paper we applied the patchy corona model to high-$z$
RQQs, and explored the model parameter space. We compared the results
to the results of the truncated disk with the hot inner flow model
studied in Paper I. Our results show that: 

\noindent
1. If the hard X-ray part of the observed high-$z$ RQQs spectra
   originates in the Comptonization process in the corona formed by
   hot clouds above an accretion disk, the clouds must be vertically
   elongated and intercept less than $\sim 10$\% of the accretion disk
   radiation. 

\noindent
2. Similarly as in the hot inner flow model, the observed X-ray
loudness and the photon index put strong constraints on the modeled
optical depth. However, the number of observables is too small to
uniquely determine the Comptonizing plasma parameters, and the grid of
solutions with different pair ($\tau$, $kT_{\rm e}$) is possible for
each object. 

\noindent
3. Both models require a similar mass of the black hole (of the order
   of $10^{10}\, M_{\odot}$), and accretion rates ($\dot{M} > 0.2\,
   \dot{M}_{\rm Edd}$). 

\noindent
4. In the patchy corona model a higher fraction of energy is allowed
to be dissipated in the hot plasma above the disk than in the hot
inner flow model.

\noindent
5. The objects in the sample with X-ray spectra characterized by the
photon index of $\Gamma \le 1.5$ can be modeled only assuming the
outflowing corona.

\noindent
6. Future high quality observational data, both in the optical/UV and X-rays
can significantly improve constraints on the model parameters and help to
discriminate between different accretion flow geometries.

\acknowledgements
We thank our referee for careful reading of the manuscript.
This research is funded in part by NASA contract NAS8-39073.  Partial
support for this work was provided by the National Aeronautics and Space
Administration through Chandra Award Number GO1-2117B and GO2-3148A issued
by the Chandra X-Ray Observatory Center, which is operated by the
Smithsonian Astrophysical Observatory for and on behalf of NASA under
contract NAS8-39073. MS and PTZ were partially supported by
PBZ-KBN-054/P03/2001, and KBN projects number 2P03D00322,
2P03D01225. MS acknowledges support from  the Smithsonian Institution
Pre-doctoral fellowship program.

\clearpage

\begin{deluxetable}{l ccccccc cccc}
\tabletypesize{\scriptsize}
\tablecaption{Modeling the SED of high-$z$ RQQs. \label{tab:mgntb0}}
\tablewidth{0pt} 
\tablehead{
\colhead{\sc Object} & \colhead{$m$\tablenotemark{a}} & \colhead{$\dot{m}$\tablenotemark{b}} &
\colhead{$f$\tablenotemark{c}} & \colhead{$\mu_{\rm s}$\tablenotemark{d}} &
\colhead{$C$\tablenotemark{e}} & \colhead{$\beta$\tablenotemark{f}} &
\colhead{$kT_{\rm e}$\tablenotemark{g}} & \colhead{$\tau$\tablenotemark{h}} &
\colhead{$\Gamma_{\rm fit}$\tablenotemark{i}} & \colhead{$\alpha_{\rm
  UV}$\tablenotemark{i}} & \colhead{$\alpha_{\rm ox}$\tablenotemark{i}}
}
\startdata
BRI0103+0032 & 1$\times$10$^{10}$ & 0.31 & 0.11 & 0.9 & 0.004 & 0 &
100 & 0.91 & 1.82 & -0.35 & 1.56\\
~& 1$\times$10$^{10}$ & 0.31 & 0.11 & 0.9 & 0.0008 & 0 & 300 & 0.22 &
1.84 & -0.35 & 1.57\\
~& 6$\times$10$^9$ & 0.6 & 0.09 & 0.9 & 0.0035 & 0 & 100 & 0.91 &
1.82 & -0.14 & 1.55\\
~& 6$\times$10$^9$ & 0.6 & 0.1 & 0.8 & 0.0035 & 0 & 300 & 0.22 &
1.82 & -0.15 & 1.54\\
~& 6$\times$10$^9$ & 0.6 & 0.09 & 0.8 & 0.001 & 0 & 100 & 0.88 &
1.84 & -0.14 & 1.54\\
~& 6$\times$10$^9$ & 0.6 & 0.09 & 0.3 & 0 & 0 & 100 & 0.64 &
2.05 & -0.14 & 1.51\\
~& 6$\times$10$^9$ & 0.63 & 0.07 & 0.5 & 0.0035 & 0.5 & 100 & 0.91 &
1.82 & -0.14 & 1.55\\
~&~&~&~&~&~&~&~&~&~&~&~\\ 
\tableline
~&~&~&~&~&~&~&~&~&~&~&~\\
PSS0248+1802 & $1{\times}10^{10}$ & 0.6 & 0.11 & 0.8 & 0.004 & 0 & 100
& 0.81 & 1.88 & -0.22 & 1.52\\
~& $1{\times}10^{10}$ & 0.60 & 0.11 & 0.8 & 0.006 & 0 & 300 & 0.20 &
1.85 & -0.23 & 1.54\\ 
~&~&~&~&~&~&~&~&~&~&~&~\\ 
\tableline
~&~&~&~&~&~&~&~&~&~&~&~\\
BRI1033-0327 & 1$\times$10$^{10}$ & 0.35 & 0.15 & 0.5 & 0.15 & 0 & 100
& 0.36 & 2.40 & -0.32 & 1.52\\
~& $1{\times}10^{10}$ & 0.35 & 0.13 & 0.1 & 0.35 & 0 & 300 & 0.04 &
2.30 & -0.33 & 1.56\\
~& $1{\times}10^{10}$ & 0.43 & 0.3 & 0 & 1 & 0 & 100 & 0.20 &
2.78 & -0.28 & 1.49\\
~& $1{\times}10^{10}$ & 0.4 & 0.153 & 0 & 1 & 0 & 300 & 0.02 &
2.45 & -0.30 & 1.58\\
~& $1{\times}10^{10}$ & 0.36 & 0.2 & 0.8 & 0.15 & -0.3 & 100 & 0.36 &
2.41 & -0.31 & 1.52\\
~&~&~&~&~&~&~&~&~&~&~&~\\ 
\tableline
~&~&~&~&~&~&~&~&~&~&~&~\\
PSS1057+4555 & $1.5{\times}10^{10}$ & 0.60 & 0.045 & 0.9 & 0.0015 & 0
& 100 & 0.9 & 1.82 & -0.29 & 1.71\\ 
~& $1.5{\times}10^{10}$ & 0.60 & 0.047 & 0.9 & 0.0030 & 0 & 300 & 0.22 &
1.82 & -0.29 & 1.72\\
~& $1.5{\times}10^{10}$ & 0.60 & 0.05 & 0.4 & 0 & 0 & 100 & 0.66 &
2.02 & -0.29 & 1.66\\
~&~&~&~&~&~&~&~&~&~&~&~\\ 
\tableline
~&~&~&~&~&~&~&~&~&~&~&~\\
PSS1317+3531 & $7{\times}10^9$ & 0.25 & 0.15 & 0.5 & 0.13 & 0 & 100 & 0.40 &
2.36 & -0.32 & 1.52\\
~& $7{\times}10^9$ & 0.23 & 0.13 & 0.1 & 0.55 & 0 & 300 & 0.03 &
2.86 & -0.34 & 1.47\\
~& $7{\times}10^9$ & 0.3 & 0.29 & 0 & 1 & 0 & 100 & 0.20 &
2.80 & -0.28 & 1.51\\
~& $7{\times}10^9$ & 0.25 & 0.18 & 0 & 1 & 0 & 300 & 0.02 &
2.41 & -0.33 & 1.55\\
~&~&~&~&~&~&~&~&~&~&~&~\\ 
\tableline
~&~&~&~&~&~&~&~&~&~&~&~\\
SDSS1030+0524 & $1{\times}10^{10}$ & 0.22 & 0.037 & 0.9 & 0.0007 & 0 & 100 &
0.95 & 1.79 & -0.42 & 1.78\\
~& $1{\times}10^{10}$ & 0.22 & 0.040 & 0.9 & 0.0010 & 0 & 300 & 0.25 &
1.78 & -0.42 & 1.79\\
~&~&~&~&~&~&~&~&~&~&~&~\\ 
\tableline
~&~&~&~&~&~&~&~&~&~&~&~\\
SDSS1204-0021 &  1$\times$10$^{10}$ & 0.3 & 0.045 & 0.9 & 0.0007 & 0 & 100
& 0.96 & 1.78 & -0.35 & 1.73\\ 
~& $1{\times}10^{10}$ & 0.2 & 0.050 & 0.9 & 0.001 & 0 & 300 & 0.27 &
1.77 & -0.35 & 1.74\\
~& $1{\times}10^{10}$ & 0.3 & 0.045 & 0.8 & 0 & 0 & 100 & 0.88 &
1.84 & -0.35 & 1.71\\
~&~&~&~&~&~&~&~&~&~&~&~\\ 
\tableline
~&~&~&~&~&~&~&~&~&~&~&~\\
PSS1443+2724 & $8{\times}10^9$ & 0.27 & 0.08 & 0.9 & 0.0035 & 0 & 100 & 0.88 &
1.83 & -0.33 & 1.61\\ 
~& $8{\times}10^9$ & 0.27 & 0.085 & 0.9 & 0.004 & 0 & 300 & 0.23 &
1.80 & -0.33 & 1.63\\ 
~& $8{\times}10^9$ & 0.27 & 0.08 & 0.6 & 0 & 0 & 100 & 0.73 &
1.95 & -0.33 & 1.59\\ 
~&~&~&~&~&~&~&~&~&~&~&~\\
\enddata

\tablenotetext{a}{Black hole mass in Solar masses, $M_{\odot}$}
\tablenotetext{b}{Accretion rate in Eddington units, $\dot{M}_{\rm Edd}$}
\tablenotetext{c}{Fraction of gravitational energy dissipated in the clouds above the disk}
\tablenotetext{d}{Cloud geometry parameter}
\tablenotetext{e}{Disk covering factor}
\tablenotetext{f}{Vertical velocity of clouds}
\tablenotetext{g}{Plasma temperature in keV}
\tablenotetext{h}{Plasma optical depth}
\tablenotetext{i}{X-ray photon index allowed to vary within the 1$\sigma$ confidence 
interval, the UV spectral index between 1450\AA\ and 2500\AA\ in the
rest frame, and the X-ray loudness (we do not list $l_{\rm UV}$ from
fits since its value was fixed at that observed)} 

\end{deluxetable}

\begin{deluxetable}{l ccccccc cccc}
\tabletypesize{\scriptsize}
\tablecaption{Modeling the SED of high-$z$ RQQs. \label{tab:mgntb05}}
\tablewidth{0pt}
\tablehead{
\colhead{\sc Object} & \colhead{$m$\tablenotemark{a}} & \colhead{$\dot{m}$\tablenotemark{b}} &
\colhead{$f$\tablenotemark{c}} & \colhead{$\mu_{\rm s}$\tablenotemark{d}} &
\colhead{$C$\tablenotemark{e}} & \colhead{$\beta$\tablenotemark{f}} &
\colhead{$kT_{\rm e}$\tablenotemark{g}} & \colhead{$\tau$\tablenotemark{h}} &
\colhead{$\Gamma_{\rm fit}$\tablenotemark{i}} & \colhead{$\alpha_{\rm
  UV}$\tablenotemark{i}} & \colhead{$\alpha_{\rm ox}$\tablenotemark{i}}
}
\startdata
BRI0241-0146 & $1{\times}10^{10}$ & 0.4 & 0.025 & 0.9 & 0.0001 & 0.5 &
100 & 1.53 & 1.53 & -0.29 & 1.86\\ 
~& $1{\times}10^{10}$ & 0.4 & 0.035 & 0.9 & 0.0001 & 0.5 & 300 & 0.51 &
1.52 & -0.29 & 1.85\\
~& $1{\times}10^{10}$ & 0.4 & 0.023 & 0.85 & 0 & 0.5 & 100 & 1.53 &
1.53 & -0.29 & 1.87\\
~&~&~&~&~&~&~&~&~&~&~&~\\
\tableline
~&~&~&~&~&~&~&~&~&~&~&~\\
BRI0401-1711 & $1{\times}10^{10}$ & 0.29 & 0.05 & 0.9 & 0.0002 & 0.5 &
100 & 1.50 & 1.53 & -0.36 & 1.75\\ 
~& $1{\times}10^{10}$ & 0.29 & 0.06 & 0.9 & 0.0003 & 0.5 & 300 & 0.48 &
1.55 & -0.36 & 1.76\\
~&~&~&~&~&~&~&~&~&~&~&~\\
\tableline
~&~&~&~&~&~&~&~&~&~&~&~\\
SDSS0836+0054 & $1{\times}10^{10}$ & 0.55 & 0.06 & 0.9 & 0.0003 & 0.5 &
100 & 1.52 & 1.53 & -0.24 & 1.70\\ 
~& $1{\times}10^{10}$ & 0.55 & 0.08 & 0.9 & 0.0004 & 0.5 & 300 & 0.50 &
1.54 & -0.24 & 1.69\\
~& $1{\times}10^{10}$ & 0.55 & 0.08 & 0.5 & 0 & 0.5 & 100 & 1.08 &
1.72 & -0.24 & 1.66\\
~&~&~&~&~&~&~&~&~&~&~&~\\
\tableline
~&~&~&~&~&~&~&~&~&~&~&~\\
SDSS1306+0356 & $1{\times}10^{10}$ & 0.25 & 0.075 & 0.9 & 0.0002 & 0.5 &
100 & 1.54 & 1.51 & -0.40 & 1.69\\ 
~& $1{\times}10^{10}$ & 0.25 & 0.095 & 0.9 & 0.0003 & 0.5 & 300 & 0.51 &
1.53 & -0.41 & 1.69\\
~&~&~&~&~&~&~&~&~&~&~&~\\
\tableline
~&~&~&~&~&~&~&~&~&~&~&~\\
SDSS1621-0042 & $1.1{\times}10^{10}$ & 0.95 & 0.06 & 0.9 & 0.0002 & 0.5 &
100 & 1.58 & 1.51 & -0.17 & 1.69\\ 
~& $1.1{\times}10^{10}$ & 0.95 & 0.08 & 0.9 & 0.0003 & 0.5 & 300 & 0.52 &
1.52 & -0.17 & 1.68\\
~&~&~&~&~&~&~&~&~&~&~&~\\
\tableline
~&~&~&~&~&~&~&~&~&~&~&~\\
BRI2212-1626 & $1{\times}10^{10}$ & 0.31 & 0.05 & 0.9 & 0.0002 & 0.5 &
100 & 1.51 & 1.53 & -0.35 & 1.75\\ 
~& $1{\times}10^{10}$ & 0.31 & 0.07 & 0.9 & 0.0003 & 0.5 & 300 & 0.49 &
1.54 & -0.35 & 1.73\\
~&~&~&~&~&~&~&~&~&~&~&~\\
\enddata

\tablenotetext{a}{Black hole mass in Solar masses, $M_{\odot}$}
\tablenotetext{b}{Accretion rate in Eddington units, $\dot{M}_{\rm Edd}$}
\tablenotetext{c}{Fraction of gravitational energy dissipated in the clouds above the disk}
\tablenotetext{d}{Cloud geometry parameter}
\tablenotetext{e}{Disk covering factor}
\tablenotetext{f}{Vertical velocity of clouds}
\tablenotetext{g}{Plasma temperature in keV}
\tablenotetext{h}{Plasma optical depth}
\tablenotetext{i}{X-ray photon index allowed to vary within the 1$\sigma$ confidence 
interval, the UV spectral index between 1450\AA\ and 2500\AA\ in the
rest frame, and the X-ray loudness (we do not list $l_{\rm UV}$ from
fits since its value was fixed at that observed)}

\end{deluxetable}

\end{document}